\documentclass[11pt]{article}
\pdfoutput=1
\usepackage{jheppub}
\usepackage{graphicx}
\usepackage{amssymb}
\usepackage{amsmath,amssymb}
\usepackage{slashed}
\usepackage{hyperref}
\usepackage{caption}
\usepackage{xcolor}
\usepackage{dsfont}
\usepackage{verbatim}
\usepackage{subfig}

\newcommand\beq{\begin{equation}}
\newcommand\eeq{\end{equation}}
\newcommand\be{\begin{equation}}
\newcommand\ee{\end{equation}}

\newcommand\varp{\varphi_{RS}}

\title{Geometries with mismatched branes}

\preprint{\today}

\author[a]{Andreas Karch,}
\author[b]{Lisa Randall}
\affiliation[a]{Department of Physics, University of Washington, Seattle, WA, 98195-1560, USA}
\affiliation[b]{Harvard University, 17 Oxford St., Cambridge, MA, 02139, USA}

\abstract{We study Randall-Sundrum two brane setups with mismatched brane tensions. For the vacuum solutions, boundary conditions  demand that the induced metric on each of the branes is either de Sitter, Anti-de Sitter, or Minkowski. For incompatible boundary conditions, the bulk metric is necessarily time-dependent.  This introduces a new class of time-dependent solutions with the potential to address cosmological issues and provide alternatives to conventional inflationary (or contracting) scenarios. We take a first step in this paper toward such solutions.
One  important finding is that the resulting solutions can be very succinctly described in terms of an effective action  involving only the induced metric on either one of the branes and the radion field.  But the full geometry cannot necessarily  be simply described with a single coordinate patch. We concentrate here on the time-dependent solutions but argue that supplemented with a brane stabilization mechanism one can potentially construct interesting cosmological models this way. This is true both with and without a brane stabilization mechanism.

}

\begin{document}
\maketitle

\section{Introduction}

Randall Sundrum (RS) brane worlds \cite{Randall:1999ee,Randall:1999vf} provide a very attractive phenomenological scenario for physics beyond the Standard Model and for cosmology. They are also of great interest to address more formal questions. Recently RS brane worlds have been employed to understand the evaporation of black holes \cite{Almheiri:2019hni,Rozali:2019day} in the context of the AdS/BCFT correspondence \cite{Karch:2000gx,Karch:2000ct,Takayanagi:2011zk}. RS branes also play a crucial role in realizing accelerating universes in string theory, both the classic KKLT construction \cite{Kachru:2003aw,Randall:2019ent} as well as in the recent proposals of \cite{Antonini:2019qkt,Banerjee:2018qey,Banerjee:2019fzz}. In particular in the latter context one question that has been raised is that of ``mismatched" branes. It is this issue we wish to address in this work.

Let us first introduce what exactly is meant by mismatched branes. The simplest Randall Sundrum geometries are solutions to Einstein gravity with a negative cosmological constant in the presence of a codimension one brane. Away from the brane the spacetime is\footnote{For simplicity we discuss here the 5d case even though the discussion goes through unchanged in general spacetime dimensions.} AdS$_5$ with curvature radius $L$. The corresponding metric is written as a warped product
\begin{equation}
\label{adapted}
ds^2 = e^{2A(z)} (ds^2_4 + dz^2)
\end{equation}
where the 4d slice is taken to be a maximally symmetric space, that is Minkowski, de Sitter (dS) or Anti-de Sitter (AdS) space of unit radius (in the latter two cases). The corresponding warp factor $e^A$ is either $L/z$, $1/\sinh(z/L)$ or $1/\sin(z/L)$ depending on the choice of slicing. Of course one can allow less symmetric spacetimes on the slice, but the maximally symmetric slicings correspond to the vacuum solutions in the presence of a brane with specific tension. For a positive tension brane the full solution corresponds to two copies of the IR part of the geometry, $z>z_*$, glued together along the brane. For a negative tension brane we keep the UV ($z<z_*$) of the geometry instead. In either case, the jump in extrinsic curvature when connecting the two halves is compensated by the brane tension. Which slicing to use is dictated by the tension of the brane. To get Minkowski space on the brane the magnitude of the brane tension $|\lambda|$ must take a critical value $\lambda_c$. For larger $|\lambda|$ one gets a de Sitter solution, for smaller $|\lambda|$ an Anti-de Sitter solution. The further away one is from the critical value, the smaller the effective 4d curvature radius $l$. On the AdS side the smallest induced radius $l$ is reached at zero tension in which case $l$ is just equal to $L$. On the dS side we can get $l$ to be arbitrarily small by increasing the magnitude of the tension.

When we dial the tension towards the critical value from either side the curvature radius on the brane diverges. The topology of the branes changes discontinuously when dialing the tension through $\lambda_c$. This is apparent from the embeddings in AdS$_5$ as we display them in figures \ref{minkbrane}, \ref{standarddesitter} and \ref{standardantidesitter}. The de Sitter brane touches the boundary of AdS$_5$ only at an instant; its boundary is spacelike. Minkowski branes have a lightlike boundary. Both of them only have a finite lifetime in terms of a global time coordinate that covers all of AdS$_5$. In contrast, the AdS brane lives forever in global coordinates and its boundary is timelike. This latter property makes the AdS brane the most interesting object for more formal applications as it allows a simple holographic interpretation in the context of the AdS/BCFT correspondence \cite{Karch:2000gx,Karch:2000ct,Takayanagi:2011zk}.

Of most interest in realistic constructions, as well as for cosmological applications, are RS setups with two branes. Most of the literature on two-brane RS setups  considers only matched branes. That is the branes are either both critical, or both undercritical, or both overcritical. Generally such setups require a positive and negative tension brane. Only in the undercritical AdS case can we have two positive tension branes in causal contact with each other. The warp factor in that case turns around and reaches two separate asymptotic regions (at $z=0$ and $z=\pi$) \cite{Karch:2000ct}. A positive tension brane can be located at each side. In the other two cases the second brane, often referred to as the IR brane, must have negative tension, but its magnitude still has to be equal to the critical value in the Minkowski case and larger than the critical value in the de Sitter case (and also larger in magnitude than that of the UV brane).
AdS allows such a possibility of a positive tension UV brane paired with a negative tension IR brane as well; it is however not essential to a two-brane scenario. The IR brane locations are arbitrary (without any additional fields) in the Minkowski case but stable locations are determined by the brane tension in all other cases.

A question that had not been seriously addressed in the literature is what happens if the second brane demands a different slicing than the first -- that is if we were to add an IR brane whose tension, in magnitude, is in a different regime compared to $\lambda_c$ than the UV brane. This is what we refer to mismatched branes.

After briefly reviewing the geometry and properties of the single brane RS setups, we will study a variety of mismatched brane geometries to highlight some of their basic features. Maybe most importantly, we confirm that all our solutions can be found using a low energy effective action that only contains Einstein gravity and a single scalar, the ``radion", whose expectation value sets the effective Newton constant. While the radion action has been written down before, it has usually not been applied to scenarios like the ones we consider, in which the worldvolume geometry has significant curvature. Our geometries can be seen as solutions to the full non-linear radion-Einstein equations. The fact that the radion has generically non-trivial couplings to the Einstein-Hilbert term is absolutely crucial to make this work. This coupling allows cancellation of much of the back-reaction of the radion dynamics. The curvature coupling introduces a new term on the right hand side of Einstein's equations proportional to the second derivative of the radion field. For the special solutions we find, this contribution cancels the contribution from the standard radion stress energy tensor. Such a cancellation will of course not happen for general radion profiles.

We furthermore observe several new features not ordinarily part of the effective theory. One is that this scenario depends on a new parameter, which is the relative creation time of the two branes. As we will see, physics can depend very strongly on this parameter. In addition such a scenario requires additional boundary conditions not usually present in the usual effective theory.

We also note that such setups generally contain time-dependence that cannot be neglected. In some cases Newton’s constant itself has strong time-dependence, showing that constructions of the type presented in \cite{Banerjee:2018qey,Banerjee:2019fzz} will need some refinement to be viable. However, we also observe that precisely the time-dependence of the solutions might suggest properties that have the potential to address outstanding cosmological issues.
In particular, one novel possibility is that the four-dimenional $M_{Pl}$ itself changes dramatically as the UV brane emerges from the boundary. Although such scenarios are generally relegated to Brans-Dicke territory, the discontinuous change in the metric associated with branes bouncing off the boundary is not readily adapted to this interpretation. While intriguing, we have not yet realized this scenario in a realistic construction though we do note our solutions naturally contain both expanding and contracting universes.

This note is organized as follows. In the next section we will review the cast of characters. We introduce the various mismatched brane configurations in section 3, starting for simplicity with the case of positive tension branes only. This, in particular, will introduce us to the feature that dS UV branes can be born again when observed from a mismatched IR brane. In section 4 we focus on born again branes with a negative tension IR brane as they have many interesting features and will also serve as the template for much of the following discussion. In section 5 we show how our solutions can also be obtained from the point of view of the low energy effective action. Maybe the most important insight to be gained from this analysis is that changing location of the observer in the higher dimensional bulk from the lower dimensional point can be accounted for by a change of frame -- that is a redefinition of what we mean by the metric. In section 6 we succinctly summarize our findings by casting them in the form of an Einstein frame action, laying the groundwork for potential future cosmological applications.

\section{The cast of characters}

In order to describe solutions with multiple mismatched branes, let us first review the cast of characters and their geometry. Since in all cases the spacetime is just (two copies) of a portion of AdS$_5$, the easiest way to depict these geometries is to indicate how the brane is embedded in AdS$_5$ together with a prescription of what part to keep. In this section we will exclusively focus on the case of {\it positive} tension branes unless explicitly mentioned otherwise. In order to understand negative tension branes one simply reverses the role of the region excised and the region kept.

For each brane, there will be a special choice of coordinates on AdS$_5$ in which the position of the brane is time independent. In this coordinate system the metric takes the form of (\ref{adapted}). We will refer to this as the adapted coordinate system corresponding to a particular brane tension regime and hence a particular 4d metric. Multiple branes with different tensions can be represented as static branes in the same adapted coordinate system (\ref{adapted}) as long as they all demand the same slicing, that is they either have to all have tension (in magnitude) larger than, equal to, or smaller than the critical value. In order to understand how these branes are embedded in AdS$_5$ it is easiest to transform these ``adapted" coordinates to standard global coordinates on AdS$_5$. Since this will be a crucial construction in understanding mismatched branes, let us be quite explicit about how this is done.

The basic tool in order to work out any coordinate change in AdS$_5$ is to use the embedding space formalism, as is for example described in detail in the classic review \cite{Aharony:1999ti}. That is, we think of AdS$_5$ as a submanifold of $\mathbb{R}^{2,4}$ defined by the $SO(2,4)$ invariant equation
\be
X_0^2 + X_5^2 - \sum_i X_i^2 = 1
\ee
where we set the AdS$_5$ curvature radius $L=1$. The index $i$ runs from 1 to 4. Different coordinate systems correspond to different parameterizations of the $X_I$ (where $I$ runs over the entire 0 to 5 range). To change from one coordinate system to another one simply equates the different parameterization of $X_I$.

Standard global coordinates on AdS$_5$ correspond to the parameterization
\begin{eqnarray}
X_0 &=& \cosh \rho \, \cos \tau \nonumber \\
X_i &=& \sinh \rho \, \Omega_i \nonumber \\
X_5 &=& \cosh \rho \, \sin \tau \label{global}
\end{eqnarray}
where $\Omega_i$ is a parameterization of a unit 3-sphere, $\sum_i \Omega_i^2 = 1$. In these coordinates the AdS$_5$ metric reads
\be
ds^2 = - \cosh^2 \rho \,\, d\tau^2 + d \rho^2 + \sinh^2 \rho \,\, d \Omega_3^2
\ee
where $d\Omega_3^2$ is the metric on the unit 3-sphere. All our diagrams will use these global coordinates in which AdS$_5$ looks like a cylinder. In particular we will display ``vertical cross sections" of AdS$_5$ corresponding to
\begin{flalign} \quad \quad \mathrm{\bf Vertical \, \, cross \, \, section:} \quad
\Omega_{1,2,3}=0, \quad \Omega_4= \pm 1 \label{sideview} && \end{flalign}
as well as  ``horizontal cross sections'' at
\begin{flalign}
\quad \quad \mathrm{\bf Horizontal \, \, cross \, \, section:} \quad
 \Omega_{1,2}=0, \quad \tau=0, \quad \Omega_3 = \cos \alpha,  \quad \Omega_4 = \sin \alpha.
\label{topview} &&
  \end{flalign}
In order to be able to draw the entire spacetime we plot a re-scaled radial coordinate $\Theta$ with
\be \cosh \rho = (\cos \Theta)^{-1}, \quad \sinh \rho = \tan \Theta \ee
which ranges from 0 to $\pi/2$. For the purposes of the vertical cross section we effectively get $\Theta$ to range all the way from $- \pi/2$ to $\pi/2$ in order to account for $\Omega_4 = \pm1$, the two poles of the $S^3$.

\subsection{The Minkowski brane}

\begin{figure}
\begin{centering}
\subfloat[Vertical cross section of a single positive tension Minkowski brane. \label{minkbrane}]
{\includegraphics[scale=0.6]{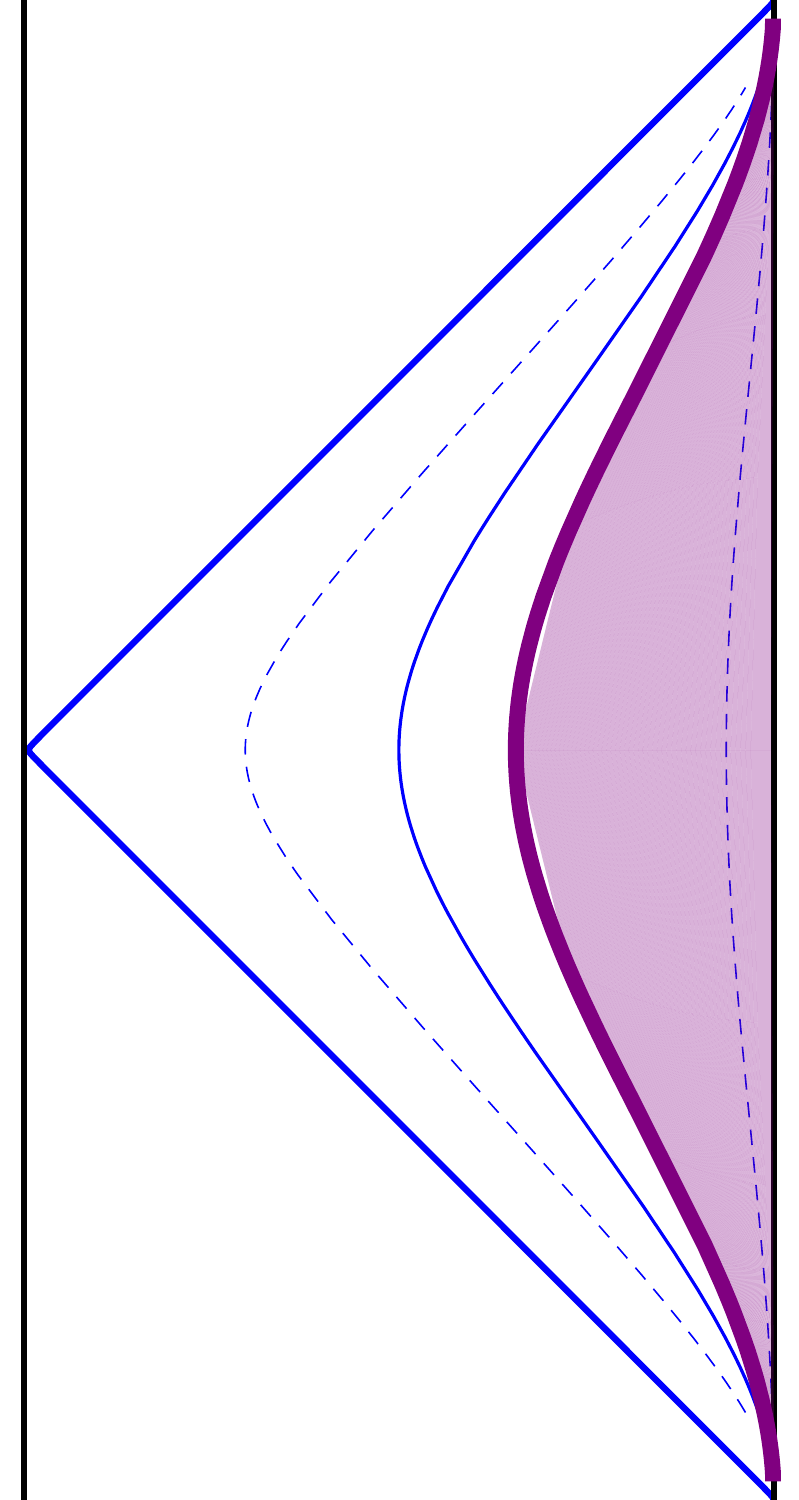}}
\hspace{2cm}
\subfloat[Standard RS setup. \label{standardrs}]
{\includegraphics[scale=0.6]{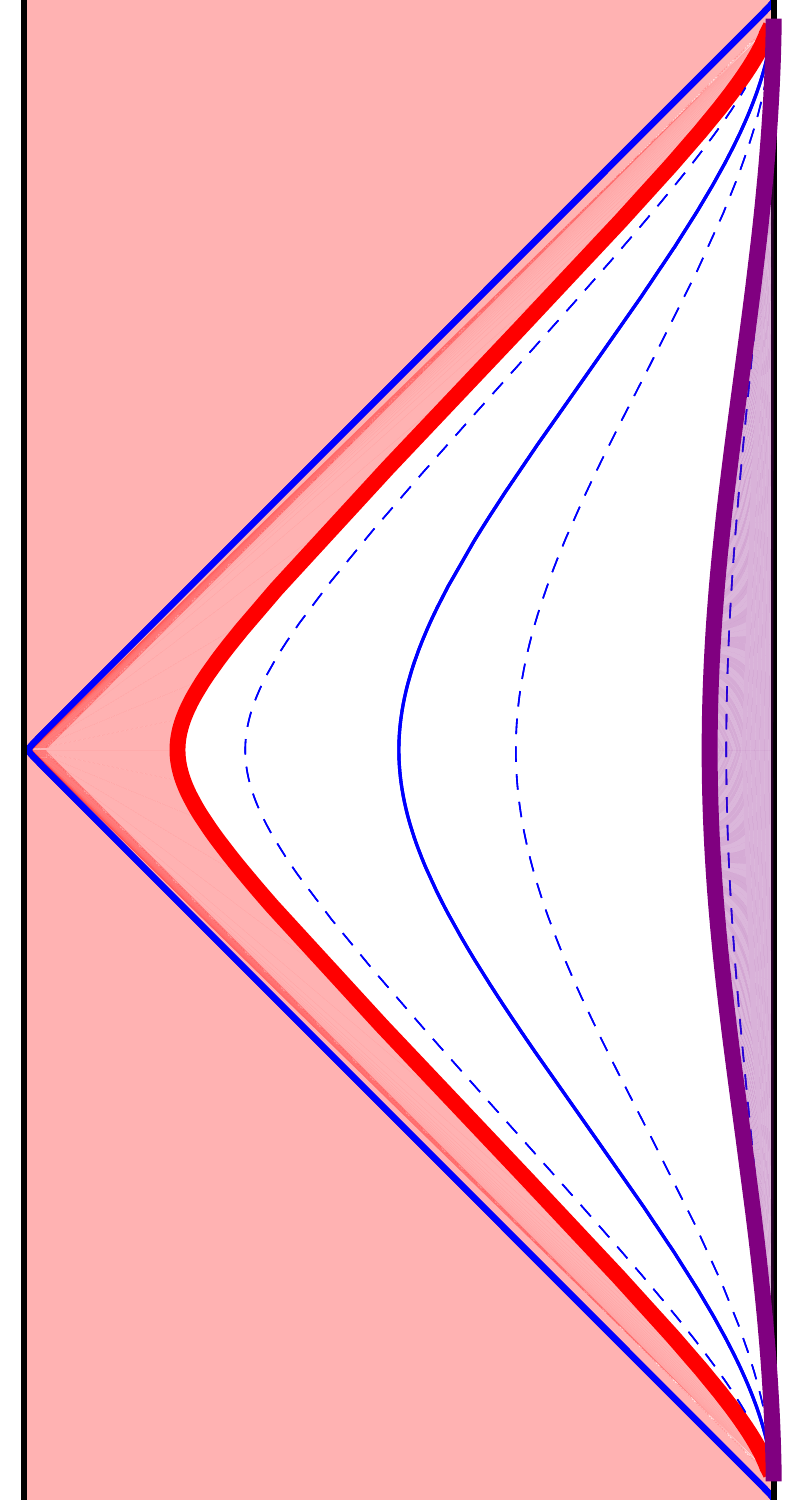}}
\caption{Vertical cross section of branes in Minkowski slicing.}
\end{centering}
\end{figure}

If we dial the brane tension to be equal to the critical value we are looking at a Minkowski brane. In this case, the adapted coordinates (\ref{adapted}) correspond to the standard Poincare patch slicing given by
\begin{eqnarray}
X_0 &=& \frac{1}{2z} \, (z^2 + x^2 - t^2 +1) \nonumber \\
X_a &=&  \frac{x}{z} \omega_a \nonumber \\
X_4 &=& \frac{1}{2z} \, (z^2 + x^2 - t^2 -1) \nonumber \\
X_5 &=&  \frac{t}{z} \label{ppatchembedding}
\end{eqnarray}
Here $a=1,2,3$ and, in analogy with the $\Omega_i$, we defined $\omega_a$ to be a parameterization of the unit 2-sphere, $\sum_a \omega_a^2 = 1$. $x$ and $\omega_a$ are the standard spherical coordinates on $\mathbb{R}^3$. In these coordinates the AdS$_5$ metric reads
\be
\label{ppatch}
ds^2 = \frac{1}{z^2} (- dt^2 + dx^2 + x^2 d \omega_2^2)
\ee
where $d\omega_2^2$ is the metric on the unit 2-sphere. The Minkowski brane is living on a slice given by $z=z_0$. Unlike all the other cases, $z_0$ is not fixed by Einstein's equations but is a modulus of the solution. The geometry of this standard RS brane is depicted\footnote{To calculate this embedding note that for the vertical cross section with $X_{1,2,3}=0$ we have $x=0$ and
\begin{eqnarray}
\tan \tau &=& \frac{X_5}{X_0} = \frac{2 t}{z^2+1 - t^2} \\
\tan \Theta &=& X_4 = \frac{1}{2z} \, (z^2  - t^2 -1)
\end{eqnarray} allowing us to translate a curve of constant $z$ into global coordinates.}
 in figure \ref{minkbrane}.
In this, and all our plots, the shaded region is excised and the full spacetime consists of two copies of the un-shaded spacetime glued together across the brane. Lines of constant $\tau$ would be horizontal lines. We explicitly displayed lines of constant $z$. The leftmost ``constant'' $z$ line, corresponding to $z = \infty$, is the horizon of the Poincare patch. No causal signal can be communicated from beyond the horizon in finite adapted time. Clearly it is inconsistent to have two positive tension Minkowski branes within a given Poincare patch: the brane at the larger $z_0$ will simply excise the entire part of the spacetime containing the second brane. We can, however, have a negative tension and a positive tension brane, retaining the spacetime between the two. This is the standard RS1 setup depicted in figure \ref{standardrs}.

One important fact to note is that the entire history of the Minkowski brane world occupies only a finite time interval in global time, which we chose to run from $-\pi$ to $\pi$. There exists a whole one parameter family of solutions for a brane with critical tension, labeled by the ``creation time" $\tau_*$ for the brane, which we here chose $\tau_* = - \pi$. Since the global AdS metric is $\tau$ translation invariant we can make this choice without loss of generality. However once we include {\it two} (or more) branes, the relative choice of $\tau_*$ matters. We could, for example, have an RS like spacetime with two positive tension Minkowski branes coexisting with different $\tau_*$ - the second brane would look like a copy of the first one translated ``up" on the cylinder. We will discuss this freedom in more detail later. Suffice it to say for now that branes appear as static only in the adapted coordinates of $(\ref{adapted})$ if they are created at the same $\tau_*$.

\subsection{The de Sitter brane}

If we increase the tension of the RS brane above its critical value, we are dealing with a de Sitter brane. For de Sitter branes, the radial position in the adapted coordinates (\ref{adapted}) is set by the tension. The larger the tension, the deeper in the IR the de Sitter brane is located and the smaller its worldvolume curvature radius becomes (corresponding to larger Hubble parameter). On the other hand, as the tension approaches its critical value from above the de Sitter brane moves closer to the boundary and its worldvolume metric becomes flatter, approaching that of Minkowski space\footnote{In terms of the adapted radial coordinate introduced in eq. (\ref{adaptedds}) the exact relation between the radial position $r_{S,0}$ of a static dS brane and its tension is \cite{DeWolfe:1999cp,Kaloper:1999sm,Kim:1999ja,Nihei:1999mt}
\beq \coth r_{S,0} = \frac{|\lambda|}{\lambda_c}.
\label{desitterposition}
\eeq}.

\begin{figure}
\begin{centering}
\subfloat[Standard de Sitter. \label{standarddesitter}]
{\includegraphics[scale=0.6]{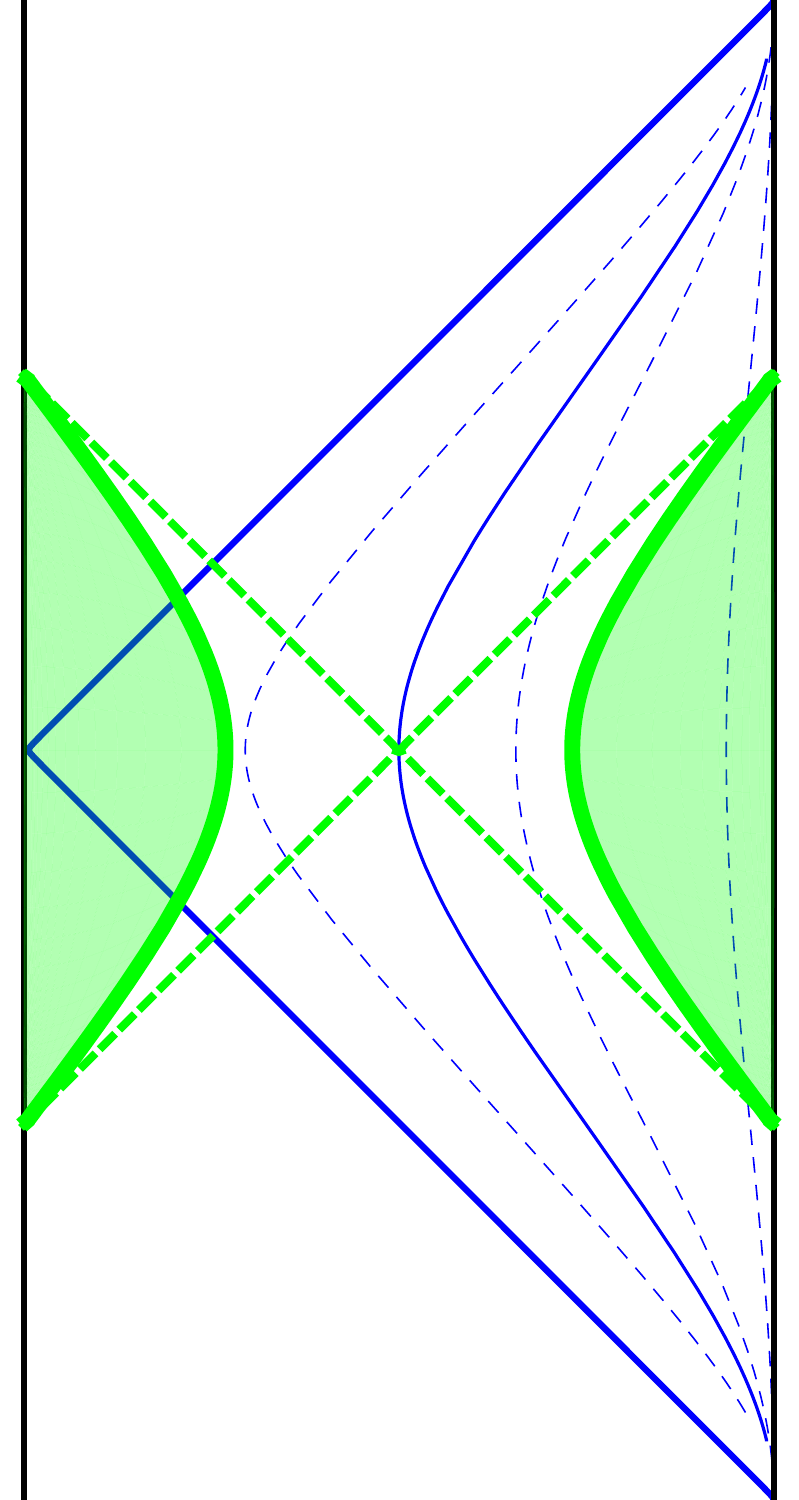}}
\hspace{2cm}
\subfloat[Alternate de Sitter. \label{alternatedesitter}]
{\includegraphics[scale=0.6]{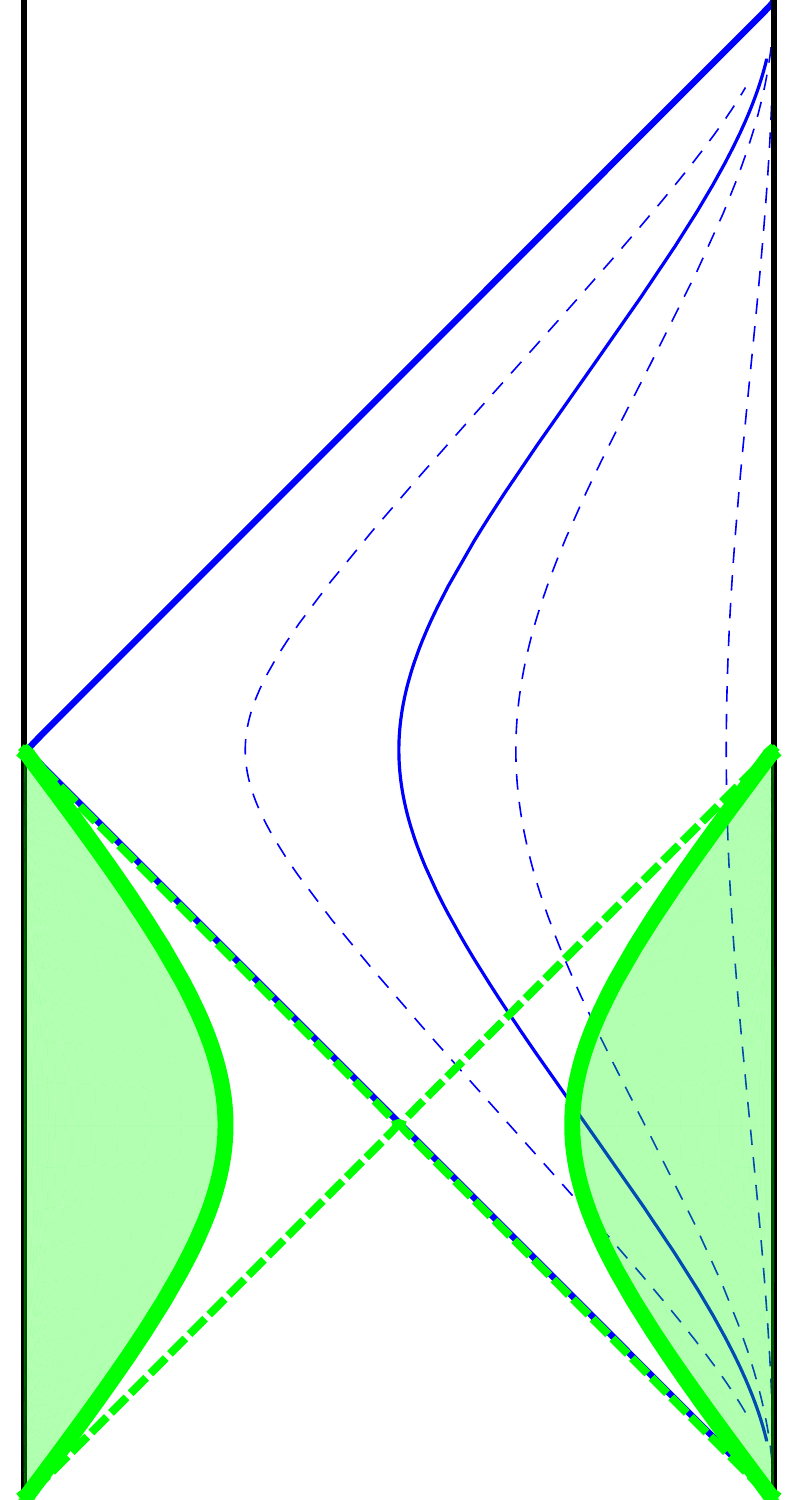}}
\caption{Vertical cross section of de Sitter branes with different $\tau_*$ together with the lines of constant Poincare $z$.
\label{desitteruniverses}}
\end{centering}
\end{figure}

For the purposes of finding the change of variables from adapted de Sitter coordinates to global coordinates we find it convenient to work with a redefined adapted radial coordinate $r_S$ in which the adapted warped product metric of (\ref{adapted}) for de Sitter now reads
\be
\label{adaptedds}
ds^2 = \sinh^2(r_S) ds^2_{dS_4} + dr_S^2.
\ee
That is, the warp factor in front of $dz^2$ got absorbed into $dr_S^2$.
To work out the change of coordinates we need to parameterize the embedding space coordinates in terms of de Sitter slices.
This is easily accomplished by noting that unit radius de Sitter$_4$ itself can be written as an embedding in $\mathbb{R}^{1,4}$ given by the equation
\be x_5^2 - x_i^2 =- 1. \ee
Correspondingly we can get adapted coordinates with dS$_4$ slices of curvature radius $\sinh(r_S)$, and hence Hubble parameter
\beq H \equiv \frac{1}{l} = \frac{1}{\sinh(r_S)} \label{hubble},
\eeq
by choosing the de Sitter embedding
\be X_0 = \cosh(r_S), \quad   X_I^2-X_5^2  = \sinh(r_S)^2 \ee
or, equivalently the alternate de Sitter embedding
\be X_5 = \cosh(r_S), \quad X_I^2- X_0^2 = \sinh(r_S)^2 . \ee
Of course, taken in isolation, de Sitter and alternate de Sitter embeddings are completely equivalent. They do, however, correspond to de Sitter universes born at a different time $\tau_*$. As such the difference becomes meaningful when we compare them to any other embedding, such as the Minkowski embedding of the last subsection, which we chose to be created at $\tau_*= -\pi$. This sets the overall origin of the $\tau$ coordinate. With this convention, the de Sitter universe is born at $\tau_*=-\pi/2$, whereas the alternate de Sitter universe is born at $\tau_*=-\pi$. The two scenarios, together with the horizons of the corresponding adapted coordinates, are displayed\footnote{For readers more familiar with representing AdS$_5$ in the Poincare coordinates of (\ref{ppatch}) we want to note that figure \ref{palternate} shows the Poincare patch embedding of a dS brane. We'll defer the details on how this plot is obtained to that latter section.} in figure \ref{desitteruniverses}. Of course any other value of $\tau_*$ can also be realized this way by setting linear combinations of $X_0$ and $X_5$ equal to $\cosh(r_S)$, but the two choices we displayed will be the ones we are mostly working with. Note that the standard de Sitter has the feature that its time coordinate has the same time reversal symmetry as the Minkowski coordinate. The alternate de Sitter universe on the other hand is natural as it is born at the same instant as the Minkowski universe.

To be explicit about the change of coordinates into these de Sitter adapted coordinates from any of the other coordinate systems discussed in this work we need to commit to a coordinate system to use on the slice. These various coordinate systems on dS$_4$ can be derived in the very same spirit as what we are doing for AdS$_5$ itself and are nicely reviewed in \cite{Spradlin:2001pw}. We will either use global dS
\be ds^2_{dS_4} = \cosh^2(\tau_S) d\Omega_3^2 - d \tau_S^2 \ee
or  flat sliced de Sitter
\be ds^2_{dS_4} = e^{2 t_S} (dx_S^2 + x_S^2 d \omega_2^2) - dt_S^2.\ee
Note that $x_S$ is the radial coordinate of a spherical coordinate system for the Minkowski space on the slice.
The corresponding parameterizations for the standard de Sitter embedding are\footnote{Note that we could have equivalently chosen $X_0 = - \cosh(r_S)$. According to (\ref{global}) a positive $X_0$ corresponds to $|\tau| < \pi/2$, which is exactly what we want for our standard dS brane that is born at $\tau_* = - \pi/2$ and terminates at $\tau=+\pi/2$. The solution with negative $X_0$ corresponds to a dS brane born at $\tau_*= + \pi/2$ which terminates at $\tau=3 \pi/2$ or one that was created at $\tau= - 3 \pi/2$ and terminates at $\tau=-\pi/2$.}
\begin{eqnarray}
X_0 &=& \cosh r_S \nonumber \\
X_i &=& \sinh r_S \, \cosh \tau_S \, \Omega_i \nonumber \\
X_5 &=& \sinh r_S \, \sinh \tau_S
\label{globaldsembedding}
\end{eqnarray}
and
\begin{eqnarray}
X_0 &=& \cosh r_S \nonumber \\
X_a &=& \sinh r_S \, x_S e^{t_S} \omega_a \nonumber \\
X_4 &=& \sinh r_S \, (\cosh t_S - \frac{x_S^2}{2} e^{t_S}) \nonumber \\
X_5 &=& \sinh r_S \, (\sinh t_S - \frac{x_S^2}{2} e^{t_S}). \label{flatdsembedding}
\end{eqnarray}
To obtain the corresponding expressions for the alternate embedding we simply exchange\footnote{To be precise, the alternate dS we are displaying here uses $X_5 = - \cosh(r_S)$. According to (\ref{global}) negative $X_5$ corresponds to negative $\tau$, so this brane has $\tau_*=-\pi$ and terminates at $\tau=0$ as described in the text. Using $X_5 = + \cosh(r_S)$ gives an alternate dS brane that is created at $\tau_*=0$ and terminates together with the Minkowski brane at $\tau=\pi$.} $X_0$ and $X_5$.
Note that it is straightforward to obtain the vertical cross section pictures of figure \ref{desitteruniverses} for a brane sitting at a constant $r_S$ from these expressions. Setting $X_{1,2,3}=0$ and using global de Sitter coordinates we see that we have
\begin{eqnarray}
\tan \tau &=& \frac{X_5}{X_0} = \tanh r_S \, \sinh \tau_S \nonumber \\
\tan \Theta &=& X_4 = \sinh r_S \, \cosh \tau_S
\end{eqnarray} for the standard de Sitter and the $X_0 \leftrightarrow X_5$ reversed expressions for the alternate de Sitter.

One very interesting aspect of the de Sitter embeddings is that they last only for half as long as a Minkowski embedding. After what appears to be a finite amount of time from the point of view of not just the global AdS$_5$ observer, but also as seen from the point of view of a Poincare slice observer, the de Sitter brane hits the boundary. In figure \ref{desitteruniverses} we let the de Sitter universe terminate once it hits the boundary. But here the boundary conditions must be imposed. For example, we could also allow it to bounce back, maybe even with a different tension: the universe would be born again. We will explore such scenarios further below when we discuss multiple branes.

Also note that the difference in lifetime also means, as already pointed out in \cite{Karch:2000ct}, that the limit of $\lambda \rightarrow \lambda_c$ is not smooth. Locally the physics of the de Sitter brane approaches the physics of the Minkowski brane as we approach the critical tension. But the global properties are discontinuous. The total lifetime of the brane jumps from $\pi$ to $2 \pi$ at the critical tension. As we will see, the limit is also discontinuous when approaching the critical value from below. In the subcritical regime the lifetime of the brane is in fact infinite.

\subsection{The Anti-de Sitter Brane}

\begin{figure}
\begin{centering}
\subfloat[Single AdS brane. ]
{\includegraphics[scale=0.4]{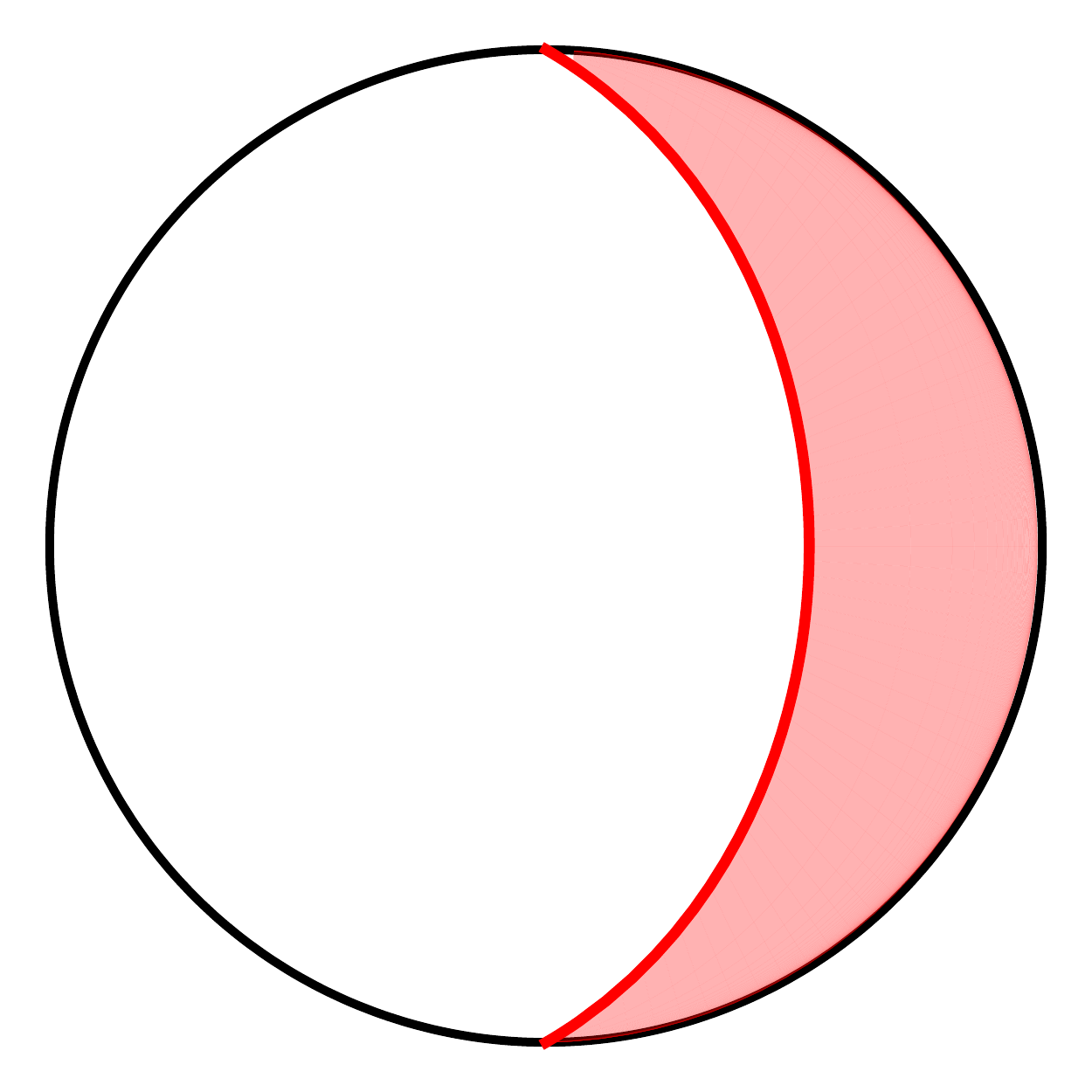} \label{adstop}}
\hspace{2cm}
\subfloat[Two positive tension AdS branes. ]
{\includegraphics[scale=0.4]{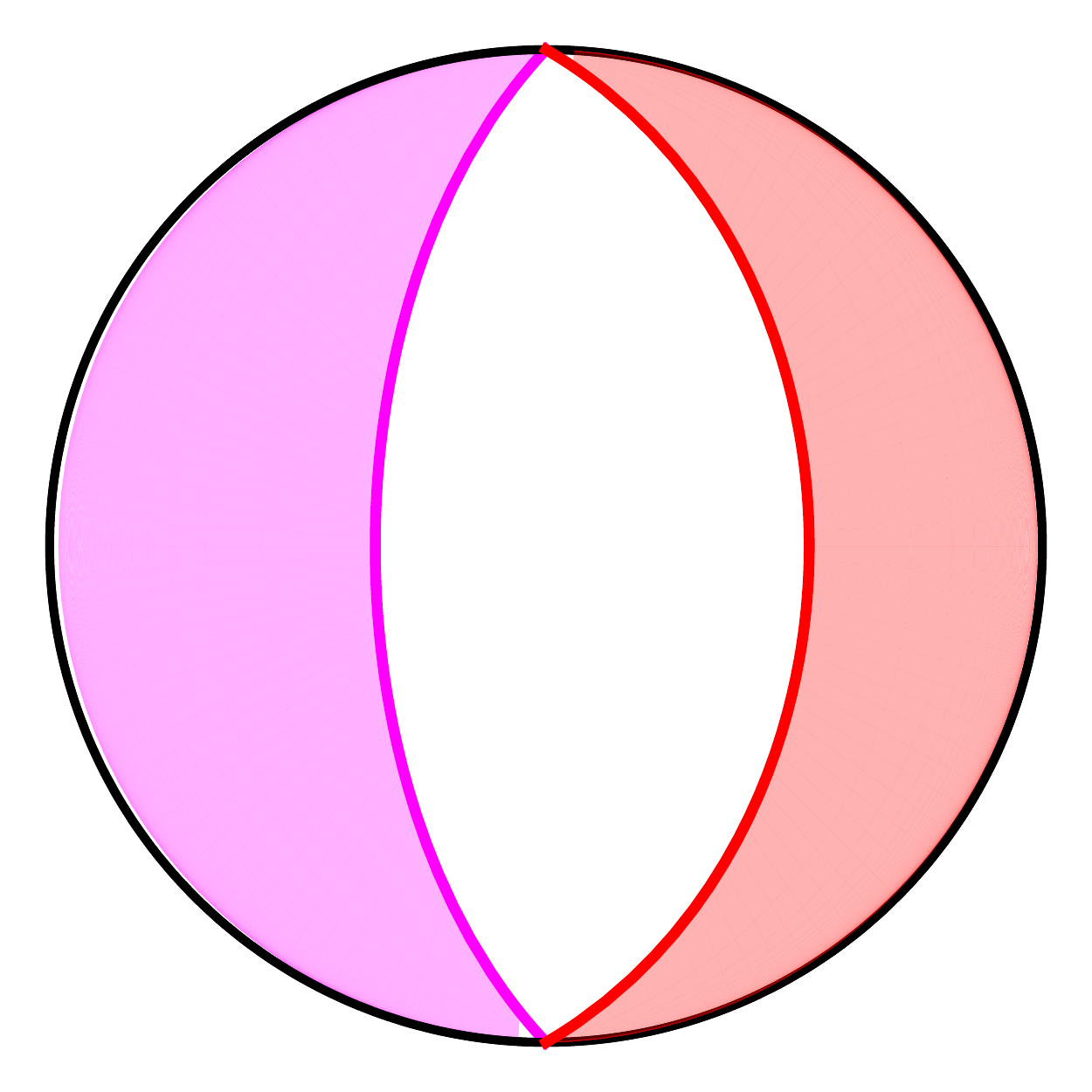}}
\caption{Horizontal cross sctions of Anti-de Sitter branes.
\label{adstopviews}}
\end{centering}
\end{figure}

Last but not least let us explore the case of the subcritical or AdS brane with $|\lambda| < \lambda_c$. This is the case relevant for the AdS/BCFT correspondence \cite{Karch:2000gx,Karch:2000ct,Takayanagi:2011zk}. In analogy with the metric (\ref{adaptedds}) we used for the de Sitter brane, the physics of the AdS brane is easiest to analyze in adapted coordinates with a redefined radial coordinate so that (\ref{adapted}) now reads:
\be
\label{adaptedads}
ds^2 = \cosh^2(r_A) ds^2_{AdS_4} + dr_A^2.
\ee
The zero tension AdS brane sits at $r_{A,0}=0$. As we increase $|\lambda|$ towards the critical value, the brane moves towards larger values\footnote{In close analogy with (\ref{desitterposition}) this time the exact relation between the static AdS brane position $r_{A,0}$ and the brane tension is given by \cite{DeWolfe:1999cp,Kaloper:1999sm,Kim:1999ja,Nihei:1999mt}
\beq
\tanh r_{A,0} = \frac{|\lambda|}{\lambda_c}
\eeq.}
 of $\cosh(r_A)$. What is special this time is that the warp factor has a turnaround at $r_A=0$ and, for any value of $|\lambda|$, there are correspondingly two allowed positions of the brane, one at positive and the other at negative $r$. Both asymptotic regions of the spacetime at $r_A \rightarrow \pm \infty$ correspond to the AdS$_5$ boundary, which appears as the union of two copies of AdS$_4$ (one from each asymptotic regions) with transparent boundary conditions: as a signal hits the boundary of the left AdS$_4$ it simply gets transmitted to the right AdS$_4$. This is just the conformally transformed statement that when using spherical coordinates for the full boundary S$^4$, which is conformal to the two copies of AdS$_4$, the equator is by no means special and signals get transmitted right through it. This can maybe most easily be seen in the horizontal cross section picture of figure \ref{adstopviews}.

\begin{figure}
\begin{centering}
\subfloat[Single AdS brane. \label{standardantidesitter}]
{\includegraphics[scale=0.6]{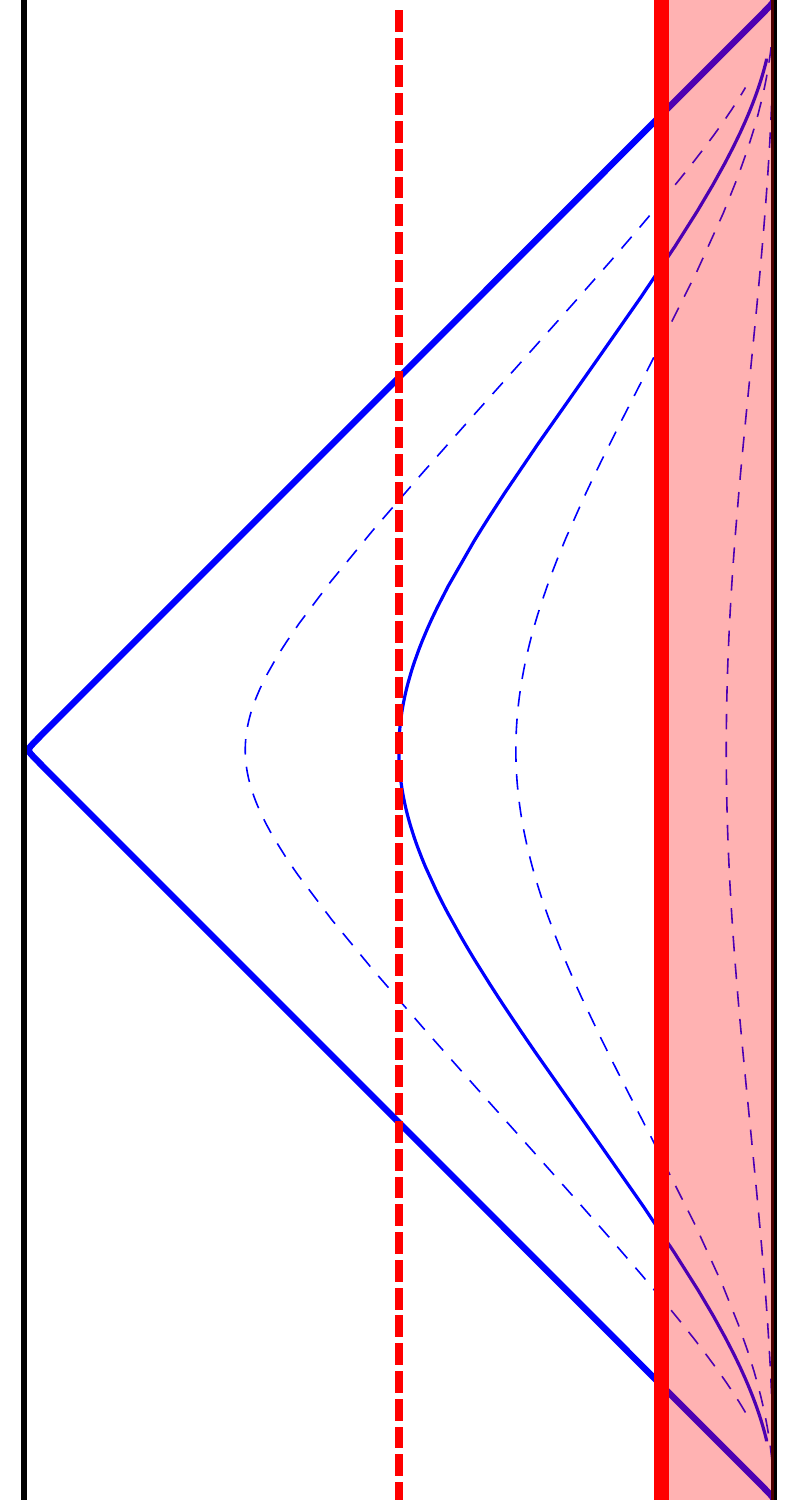}}
\hspace{2cm}
\subfloat[Two positive tension AdS branes. \label{alternateantidesitter}]
{\includegraphics[scale=0.6]{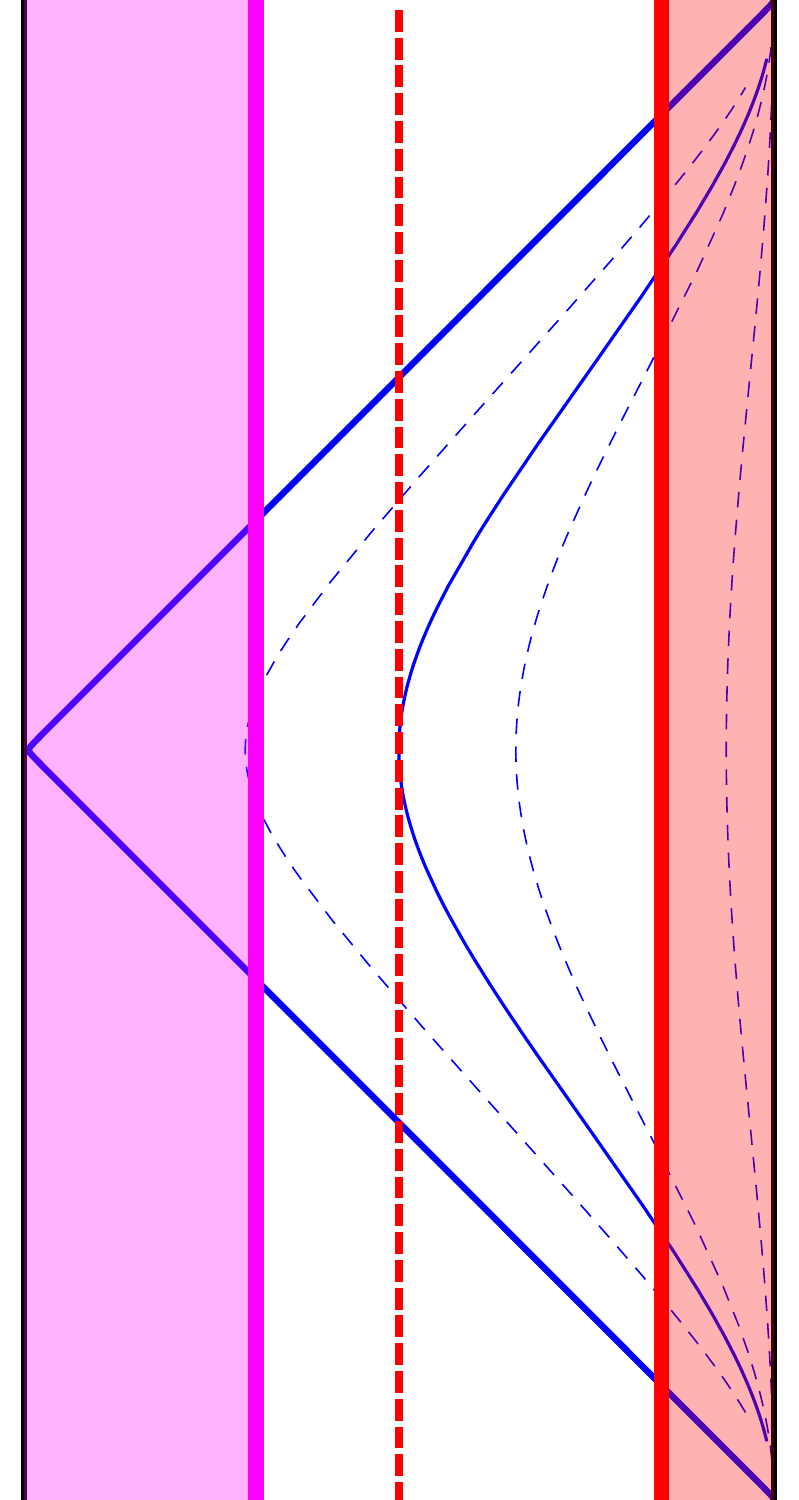}}
\caption{Vertical cross sections of Anti-de Sitter branes.
\label{antidesitteruniverses}}
\end{centering}
\end{figure}

To map the adapted AdS coordinates of (\ref{adaptedads}) to global coordinates we once again need to find the corresponding parameterization of the embedding space coordinates, this time in terms of AdS$_4$ slices. As in the dS case, this is easily accomplished by noting that if we chose
\beq
X_5^2 + X_0^2 - X_1^2 - X_2^2 -X_3^2 = \cosh^2(r_A), \quad X_4 = \sinh(r_A)
\eeq
we automatically get a foliation with an AdS$_4$ of curvature radius $\cosh(r_A)$ on each slice. The geometry of this brane is most easily visualized if we chose a global coordinate system for the AdS$_4$ on the slice as well so that its metric reads
\beq
ds^2_{AdS_4} = - \cosh^2\rho_A  \,\, d \tau_A^2 + d \rho_A^2 + \sinh^2 \rho_A \,\, d \omega_2^2.
\eeq
In terms of these adapted coordinates the embedding $X_I$ read
\begin{eqnarray}
X_0 &=& \cosh r_A \, \cosh \rho_A \,\cos \tau_A \nonumber \\
X_a &=& \cosh r_A \, \sinh \rho_A \, \omega_a \nonumber \\
X_4 &=& \sinh r_A \nonumber \\
X_5 &=& \cosh r_A \, \cosh \rho_A \, \sin \tau_A .
\label{adspara}
\end{eqnarray}
Here we used the standard parameterization of global AdS$_4$ in terms of $\rho_A$, $\tau_A$ and $\omega_a$.
This makes it clear that the vertical cross section of the AdS branes is really simple. We have
\begin{eqnarray} \tan \tau &=& \frac{X_5}{X_0}= \tan \tau_A \quad \Rightarrow \quad \tau=\tau_A \\
\tan \Theta &=& X_4 = \sinh r_A .
\end{eqnarray}
That is, an AdS brane located at a fixed value of $r_A$ in the vertical cross section simply sits at a fixed $\Theta$ in global coordinates. The AdS brane does not have a creation time $\tau_*$ and no end time. We display a vertical cross section of an AdS brane in figure \ref{standardantidesitter}. The dashed line going vertically down the middle is the location of the turnaround point.

As mentioned above, the presence of the turnaround point allows us in the AdS case, and only in the AdS case, to include a second positive tension brane consistent with the same slicing by placing it on the other side of the turnaround. This is displayed in figure \ref{alternateantidesitter}. Placing both branes on the same side of the turnaround faces the same obstacles one encountered in the de Sitter and Minkowski case: the smaller tension brane excises the region of space that would have contained the higher tension brane to begin with.

Given that the vertical cross section of the Anti de-Sitter branes does not display its interesting geometry, it is useful to also work out the horizontal cross section in this case, which highlights the interpretation in terms of the AdS/BCFT correspondence. Recall that for the horizontal cross section we set $\tau=0$, $\Omega_3 = \cos \alpha$ and $\Omega_4 = \sin \alpha$ with $\Omega_{1,2} = 0$. In our AdS adapted coordinates this is achieved by setting $\tau_A=\omega_{1,2}=0$ and $\omega_3=1$, which follows from comparing (\ref{global}) and (\ref{adspara}). Matching between global and AdS adapted coordinates for the horizontal cross section then yields
\beq
\tan \Theta = \frac{X_4}{\sin \alpha} = \frac{\sinh r_A}{\sin \alpha}.
\eeq
This makes it easy to display the horizontal cross section $\Theta(\alpha)$ for an AdS brane, which is located at a fixed value of $r_A$. We display the horizontal cross section for a single AdS brane and for two positive tension AdS branes with different tensions, separated by the turnaround point, in figure \ref{adstopviews}.

\subsection{Euclidean Solutions}
\begin{figure}
\begin{centering}
\subfloat[Horizontal cross section of Minkowski brane. ]
{\includegraphics[scale=0.4]{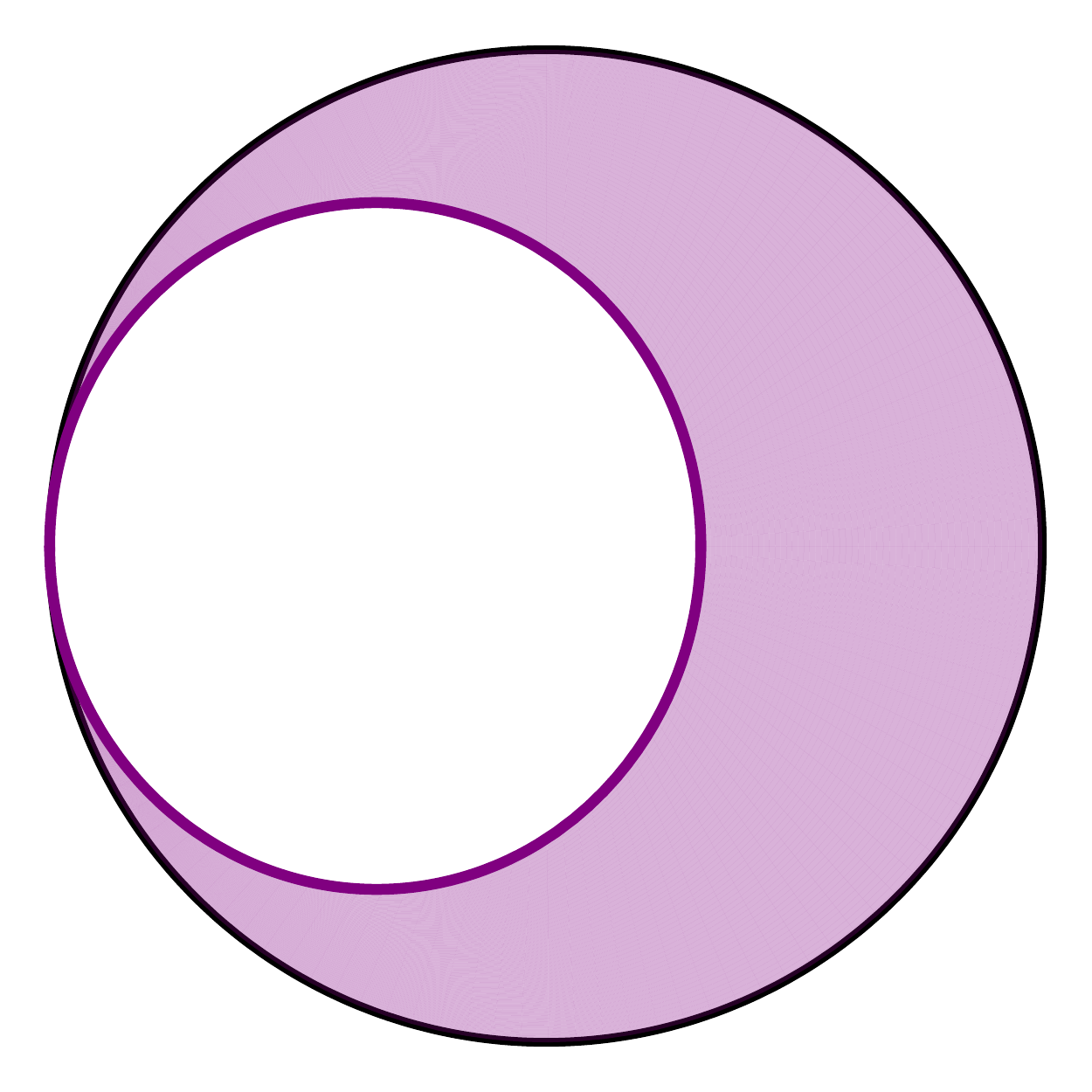} \label{minktopview}}
\hspace{2cm}
\subfloat[Horizontal cross section of de Sitter brane. ]
{\includegraphics[scale=0.4]{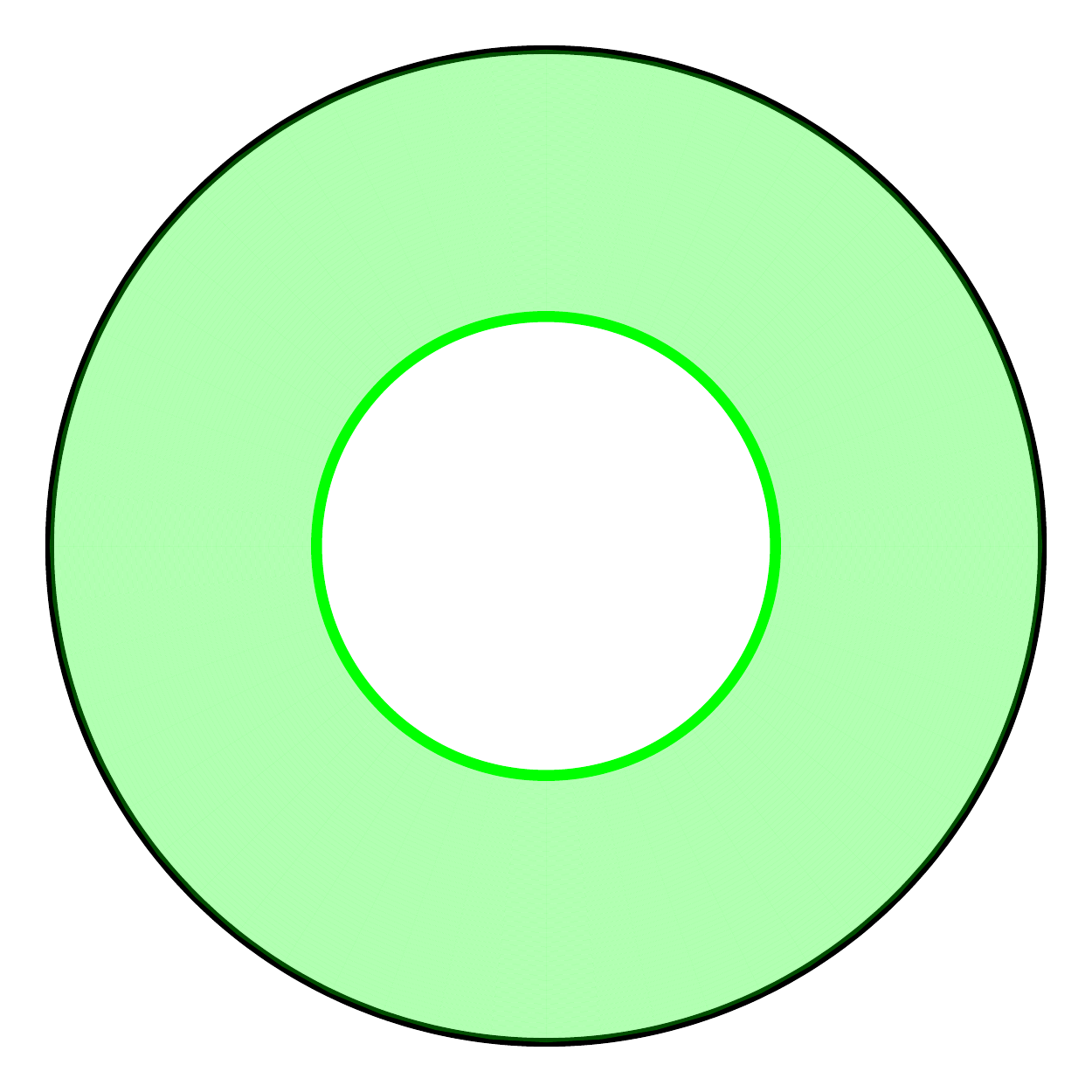}\label{dstopview}}
\caption{Euclidean embeddings of critical and supercritical brane. This is equivalent to the horizontal cross sections of Minkowski and de Sitter branes.
\label{othertopviews}}
\end{centering}
\end{figure}

For many applications, most notably in the context of AdS/BCFT, it is also useful to study these branes in a Euclidean setting. This is also required if one wants to understand how the geometry can be created by an instanton. For reasons that will become clear momentarily, we will study Euclidean RS spacetimes in 4 instead of 5 spacetime dimensions. As before, the formulas for the coordinate changes hold in any dimension to begin with so this is simply to be able to cut down on the number of indices and can be done without loss of generality. In Euclidean signature AdS$_4$ has the geometry of the interior of a ball with an $S^3$ boundary. The metric is given by
\beq
ds^2 = d \rho^2 + \sinh^2 \rho  \, \, d \Omega_3^2.
\eeq
Note that this is identical to the spatial part of eq. (\ref{global}). In fact, we can use this to derive the Euclidean embeddings from the Lorentzian counterparts in one dimension up. We simply set $\tau=0$ in all the geometries discussed so far, that is we focus on the ``horizontal cross section". For the subcritical branes with $|\lambda| < \lambda_c$ we already worked out the horizontal cross sections in the previous subsection in figure \ref{adstop}. Once again, it is obvious from these pictures that these branes are the ones that play a role in AdS/BCFT. The horizontal cross sections for the supercritical $|\lambda| >\lambda_c$ and critical $|\lambda|=\lambda_c$ branes are displayed in figure \ref{othertopviews}.

To obtain these plots we use the embeddings (\ref{globaldsembedding}) in the dS case and (\ref{ppatchembedding}) in the Minkowski case. Comparing to (\ref{global}) we see that for dS the horizontal cross sections corresponds to $r_S=\rho$, and so lines of constant $r_S$ are simply spheres at constant $\rho$. For the Minkowski brane the horizontal cross section is more complicated. At $\tau=0$ we can write
\be \cos \Theta = \frac{1}{X_0} = \frac{2z}{z^2 + x^2 +1}, \quad \tan \alpha = \frac{X_4}{X_3}
= \frac{z^2 +x^2-1}{2x} .\ee
This gives a parameterization of the $\Theta(\alpha)$ curve in terms of $x$.

Note that while the critical brane touches the boundary at exactly one point, the supercritical brane does not touch the boundary at all. Once again, the transition from the subcritical brane, which cuts the boundary in half, to the supercritical brane, which doesn't touch the boundary at all, is not smooth globally. Locally the two become indistinguishable as the tension approaches the critical value.

\section{Mix and Match}

Having carefully laid out the known brane constructions for the dS, Minkowski and AdS brane in the previous subsection together with coordinate changes that get us from each of them to global coordinates, it is now straightforward to find solutions corresponding to mismatched branes. Recall that for an RS brane the geometry is unaffected by the brane except at its location. The brane simply includes a local instruction of how to cut and paste spacetime together. Furthermore, in the full 5d theory with no additional matter, the matching across one brane is not affected by the presence of a second brane, so we can freely combine the basic building blocks from the last section. Here we assume that the bulk solution is locally AdS$_5$. That is, we have no matter present and have not turned on a finite temperature corresponding to radiation on the brane. Otherwise the branes communicate by modifying the bulk solution.

In many of our figures branes appear to overlap. One should keep in mind that the apparent brane overlaps are not real but are just a consequence of plotting 2d cross sections. The branes do however intersect and one may wonder what is happening at these intersections. In our action we do not include any terms that would allow branes to interact, even if they are close. Each brane is minimizing its worldvolume and is curving spacetime around it with its tension. But we do not include any direct cross-talk between the branes. Of course one would expect that this prescription breaks down near the points where branes intersect. Usually new light degrees of freedom are localized at such brane intersection. In string theory these could be strings stretching from one brane to the next. So while, as they stand, the solutions we describe are complete solutions to Einstein equations with two decoupled branes, we would expect our description to break down near the locus of brane intersections.

If we have two branes in the same tension regime (meaning subcritical, critical or supercritical), we can put them at two different radial positions in the adapted metric (\ref{adapted}). This is what we referred to as ``matched" brane tensions. The tensions do not have to be the same, but have to have the same relation to $\lambda_c$. For a dS or an AdS brane the location is determined by the tension, whereas for the Minkowski brane it was a free parameter to begin with. As noted in the previous section, only in the AdS case can we have two positive tension branes sitting at a time-independent position when viewed from the point of view of the adapted coordinate system.

For mismatched brane tensions, we can still combine the basic ingredients from the last section. Each brane independently solves the jump equation irrespective of the presence of the second brane. The price is that the brane positions can no longer both be static in any coordinate system. We will describe several examples of such mismatched pairs in the next few subsections, together with a few important lessons to be learned from them.

We will, for concreteness, focus on brane worlds with positive tension only in this entire section. As we will see, in the mismatched brane case this is not nearly as restrictive as in the matched case. If one wants to take one of the branes to have negative tension, one is again instructed to interchange which part of spacetime is cut out and which part is kept due to that brane.

We will start the discussion with the case where one of the branes is an AdS brane. This introduces the basic mix and match procedure.
We will then describe positive tension pairs including dS branes, which will exhibit two interesting new features: the setup crucially depends on a new parameter, the time of birth of the universe, and entire universes can be born again at a finite time when viewed from an observer sitting on a different brane.

In general the brane separation in all these mixed brane scenarios will depend on both space and time. There exists however one special coordinate system in which AdS, dS and Minkowski branes (with specially tuned creation times $\tau_*$) can all coexist with their separation only depending on time. While interesting that such a choice of coordinates exists, we have not found any practical use for it so far and hence relegated it to appendix
\ref{hyperbolic}.

\subsection{Pairs including an AdS brane}

\begin{figure}
\begin{centering}
\subfloat[Minkowski and AdS brane.]
{\includegraphics[scale=0.6]{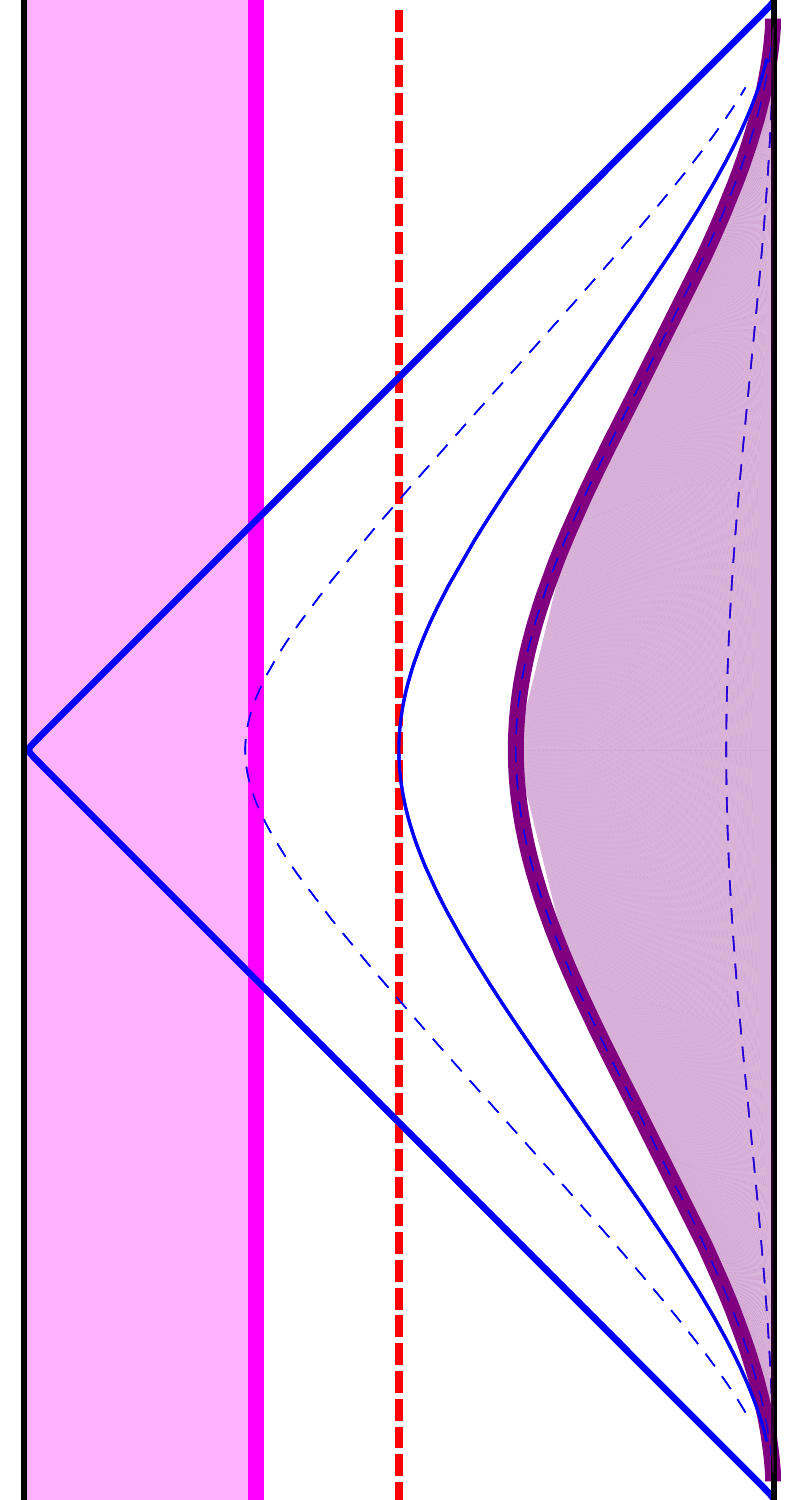} }
\hspace{2cm}
\subfloat[dS and AdS brane. ]
{\includegraphics[scale=0.6]{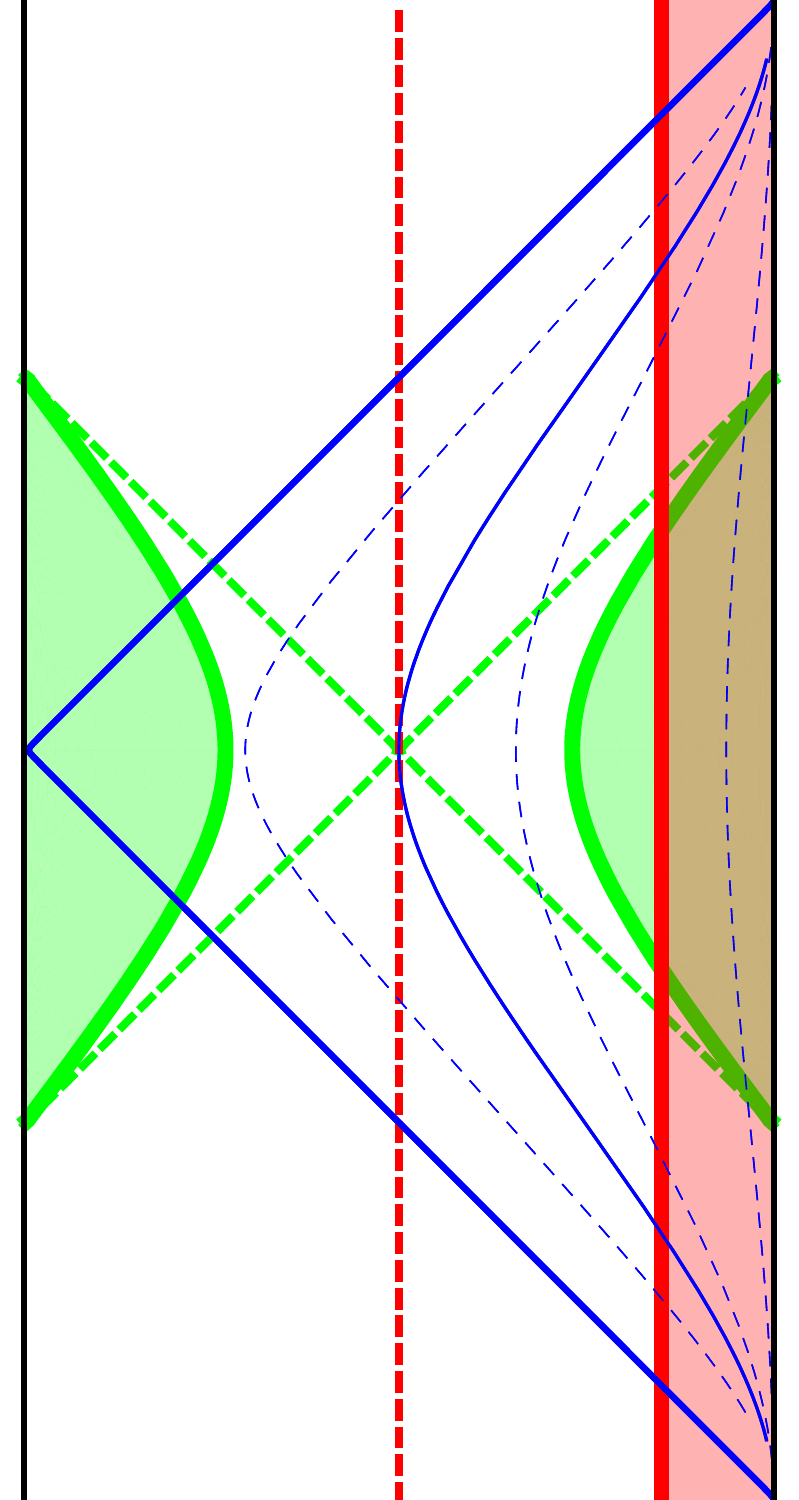}}
\caption{Vertical cross sections of mismatched branes involving an AdS brane.
\label{mixedads}}
\end{centering}
\end{figure}

Vertical cross sections of spacetimes with an AdS/Minkowski pair and an AdS/dS pair are displayed in figure \ref{mixedads}.

A few general lessons can be extracted from this. First note that in both examples we mixed two positive tension branes. This is much easier in the mismatched case. The turnaround of the AdS adapted coordinate system allows us to put either a dS or a Minkowski brane with positive tension on the other side of the turnaround. The limitations of the purely Minkowski or purely dS setup do not constrain the mismatched system.

A second obvious fact to note is that the brane separation, the ``expectation value of the radion", in both cases is time dependent, so is the volume of the warped spacetime. A time dependent radion means that we have a time dependent Newton constant. This is in general problematic for any real application. Unless one is very careful about how one exactly sets up the branes, one will in general not get a viable cosmology. We will comment more on that in the next section. This is certainly a problem to be addressed in scenarios such as the one laid out in \cite{Banerjee:2018qey,Banerjee:2019fzz} which want to get a dS cosmology out of a mismatched brane system. Last but not least the construction also makes it obvious that any mechanism that effectively stabilizes the radion, as is implied e.g. in the KKLT construction \cite{Kachru:2003aw}, has to do so by leading to matched branes. This can be accomplished by either directly affecting the brane tensions, or by affecting the bulk cosmological constant (which sets $\lambda_c$). For a detailed discussion of the viability of KKLT and the insights one can glimpse from rephrasing it in the language of an RS construction see \cite{Randall:2019ent}.

Note that one would in general expect that the brane separation is not just time but also space dependent. Not only is Newton's constant time dependent, it also depends on the location in the universe. In some special cases of mismatched branes the radion only depends on time, not space. But there are also examples where this is not possible. We will encounter examples of both kinds below.

\subsection{Born again: pairs including a dS brane}

\begin{figure}
\begin{centering}
\subfloat[Minkowski and dS brane.]
{\includegraphics[scale=0.6]{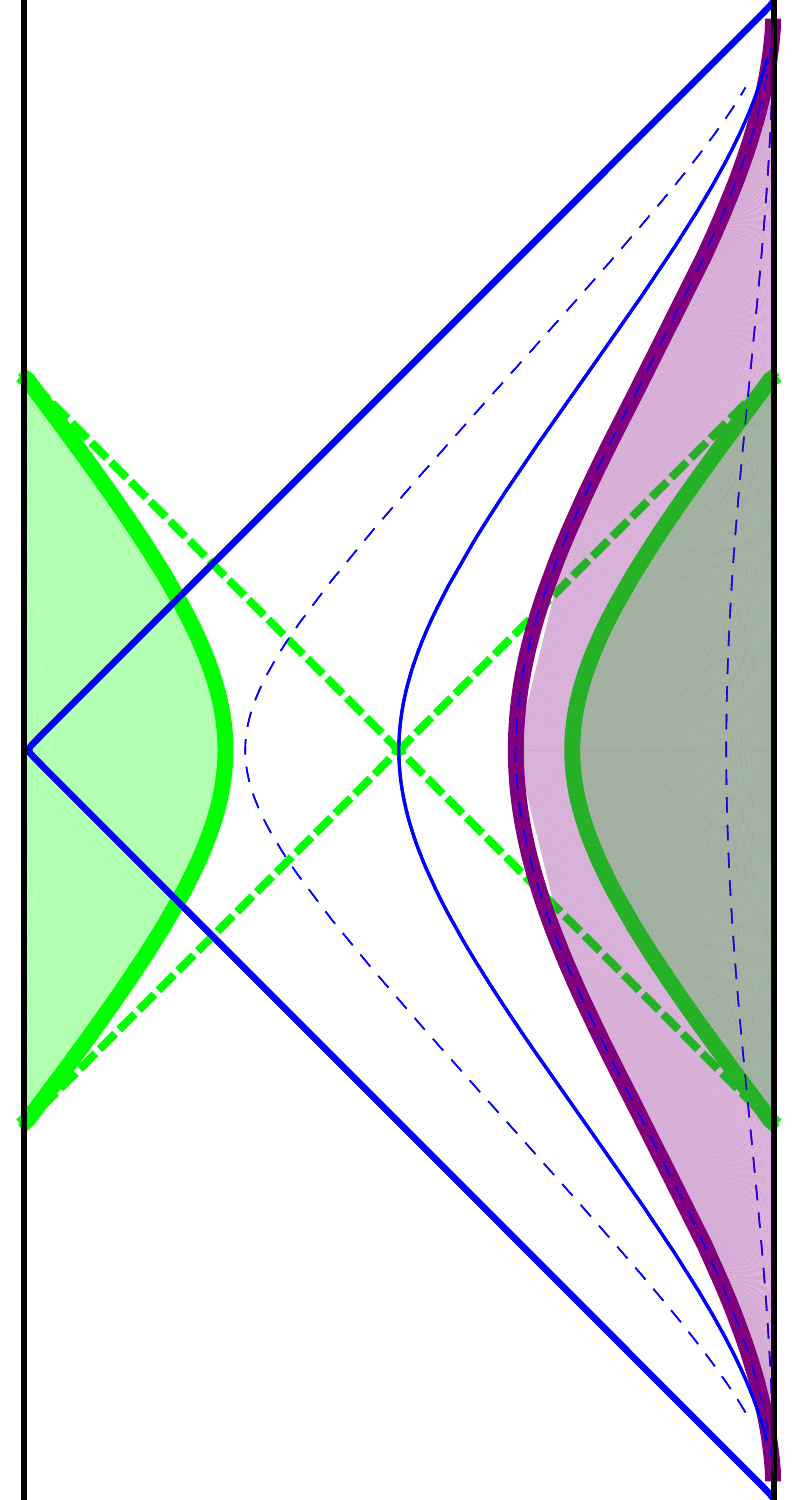} \label{minkds} }
\hspace{2cm}
\subfloat[Minkowski and two dS branes. ]
{\includegraphics[scale=0.6]{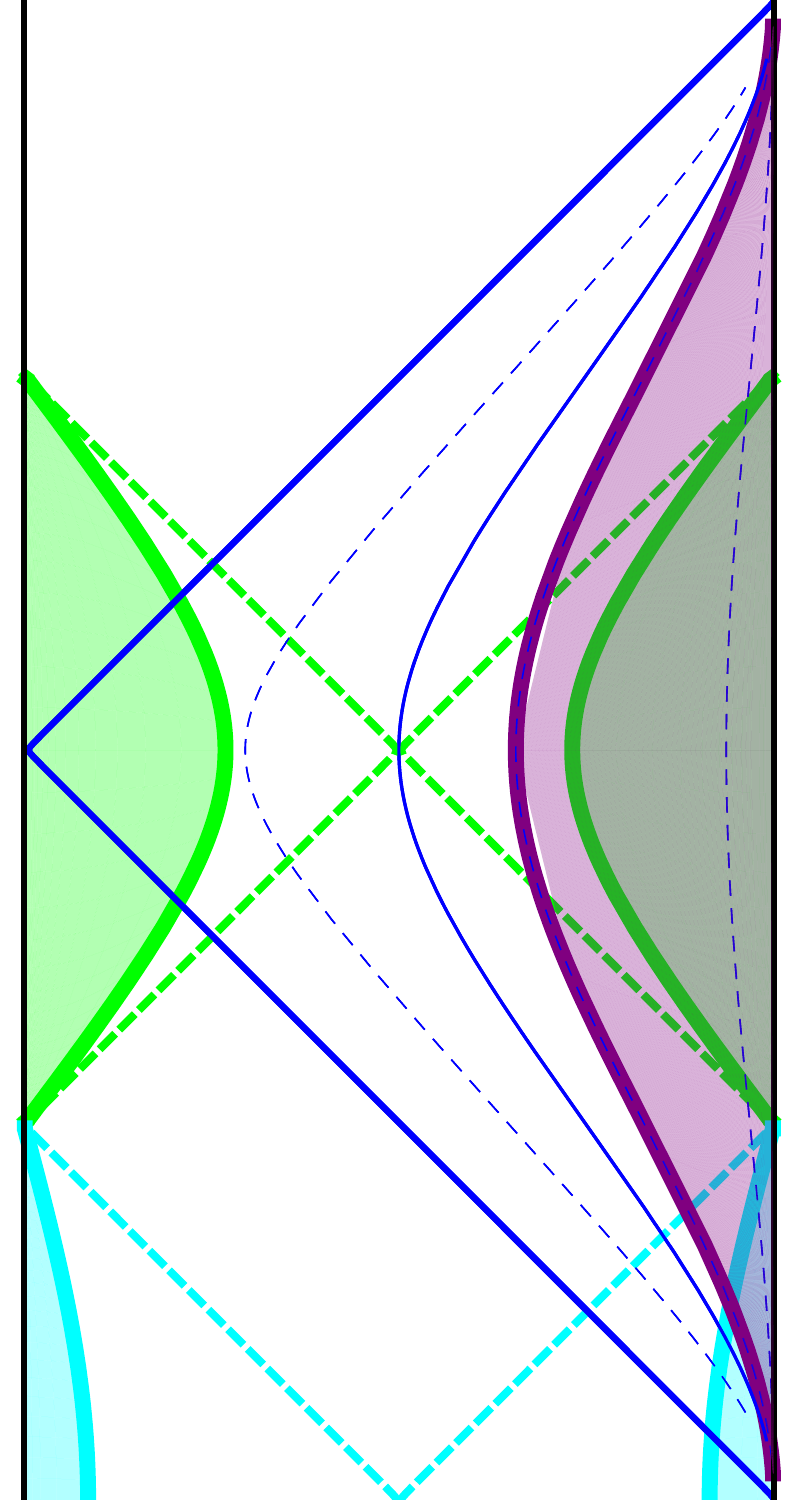} \label{minkds2}}
\caption{Vertical cross sections of mismatched branes involving dS branes.
\label{mixedds}}
\end{centering}
\end{figure}

Vertical cross sections of spacetimes with a dS/Minkowski pair as well a an dS/dS/Minkowski triple are displayed in figure \ref{mixedds}. Once again all branes in this figure have positive tension.

In addition to the general lessons extracted from the mismatched branes involving an AdS brane, there is something qualitatively new happening in the setups mixing Minkowski and dS branes. As explained in the previous section, both dS and Minkowski branes only have a finite lifetime from the point of view of the global AdS$_5$ time. Correspondingly, they are characterized by a creation time $\tau_*$. As before, we fix the creation time for the Minkowski brane to be $\tau_*=-\pi$. This leaves the creation time of the dS brane as a free parameter. In figure \ref{minkds} we display a Minkowski brane together with a dS brane created at $\tau_*=-\pi/2$. We can move the dS brane up and down relative to the Minkowski brane by adjusting its $\tau_*$. Since the universe is created at the boundary of AdS$_5$ $\tau_*$ should presumably be viewed as a parameter set by boundary conditions.

While the freedom in choosing $\tau_*$ gives us a much richer possibility for constructing 2-brane brane worlds, this new freedom does not give rise to any new scenarios with static radion. In order to be able to include two branes with time independent positions in the same adapted coordinate system (\ref{adapted}) they do not just need matched tensions, they also need to be created at the same $\tau_*$.

In figure \ref{minkds2} we added a third brane, another dS brane but this time created at $\tau=-3 \pi/2$ and with a smaller tension. Since the lifetime of the dS brane is $\pi$, the two dS brane nicely meet at the boundary: the second dS universe starts when the first one ends. This is what we call a ``born again" brane. From the point of view of the dS observer on the first dS brane, this rebirth happens at $t=+\infty$, that is after the entire history of the dS universe. What makes these born again scenarios interesting is that the rebirth is happening at a finite time when viewed from the Minkowski observer's point of view and so we can explore the physics of witnessing the rebirth of a different universe in finite time.

As an aside, note that one more thing we can learn from figure \ref{minkds2} is that using the freedom in the creation time of the dS universes we can even construct brane worlds with two positive tension dS branes after all. Let us indicate the creation time as a subscript. We see from \ref{minkds2} that a brane world with a dS$_{\tau_* = - \pi/2}$ brane and a dS$_{\tau_* = - 3 \pi/2}$ brane all by itself is a perfectly respectable GR solution. In this case the branes are completely outside each others horizon and so an observer on either one of the branes would not notice the other brane in this special case. Only an observer using global time, or an observer on the Minkowski brane (which after all had a lifetime of $2 \pi$) in the full setup of \ref{minkds2} involving three branes, would be able to see both. But this is special to the case of having the creation time of the two dS branes being separated exactly by $\pi$, the lifetime of the dS universe. Clearly a brane world with two positive tension dS branes with creation times separated by less than $\pi$ is perfectly consistent and has the branes in causal contact. In fact, they intersect.

\section{Born again branes}

One very interesting class of mismatched RS setups involves dS UV-branes. For the purposes of this section we will return to the basic RS1 type setup with a negative tension IR brane and a positive tension UV brane. The only difference is that we slightly detune the UV brane to be supercritical. There are several interesting aspects to this setup. First of all, since the lifetime of the dS universe is half of that of the Minkowski universe, we do have strong dependence on the creation time $\tau_*$ of the de Sitter universe. Second, since the dS brane hits the boundary at a finite Minkowski time, we do need to specify what happens with the brane at this instant. For this section, we chose to always have de Sitter universes born again: they bounce back from the boundary whenever they hit it. From the point of view of the Minkowski brane this means we are describing a very curious event: there exists an instant in time where Newton's constant vanishes. These born again branes describe very interesting cosmological scenarios, albeit apparently quite different from what we need to describe the universe we live in. We'll describe some of their most interesting features below. For now we will simply lay out the geometry of these reborn universes as they will prove to be very useful in order to construct and understand the effective action describing mismatched branes.

\begin{figure}
\begin{centering}
\subfloat[Reborn Universe \label{standardbouncing}]
{\includegraphics[scale=0.6]{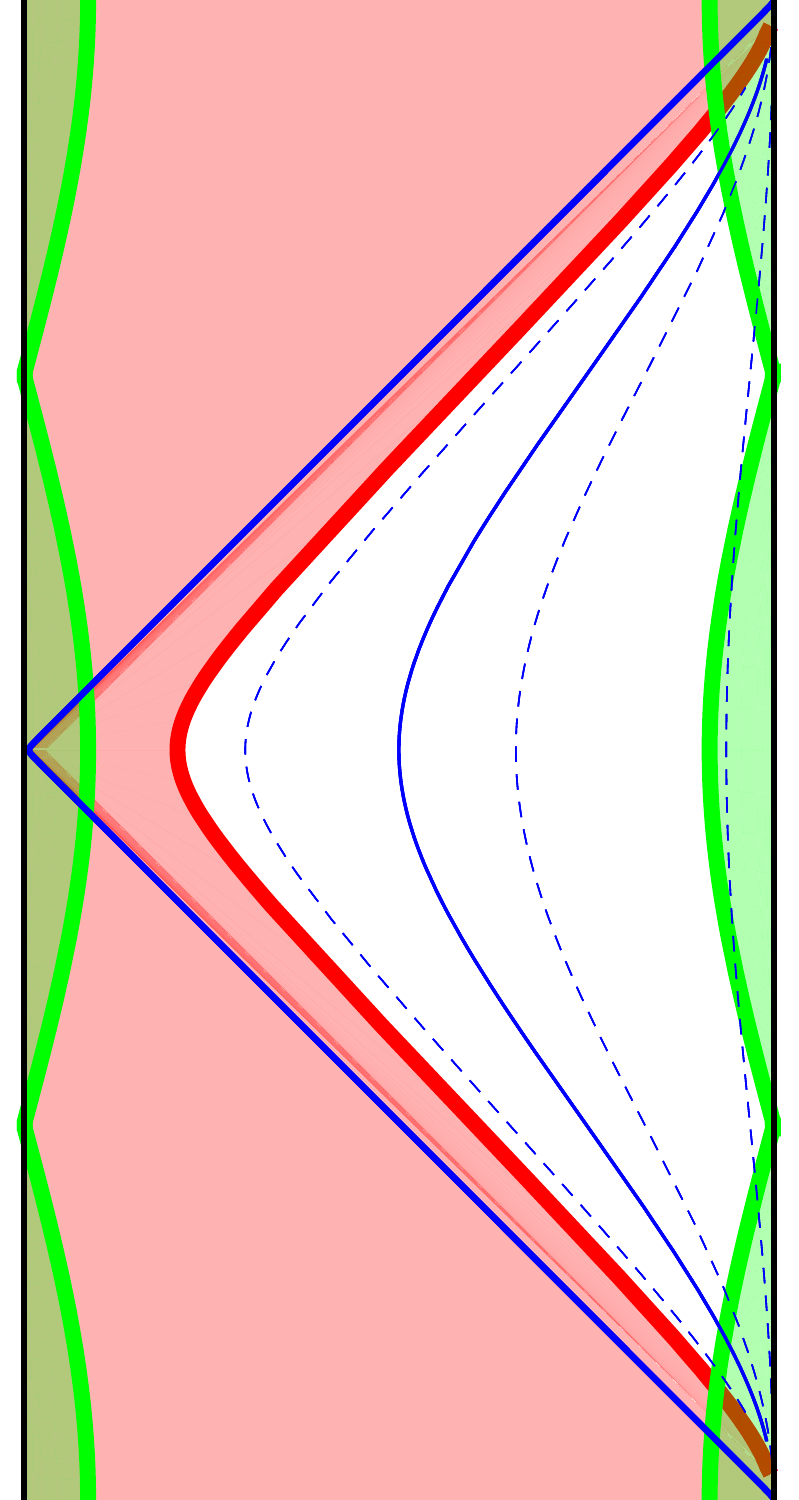}}
\hspace{1cm}
\subfloat[Alternate reborn scenario. \label{alternatebouncing}]
{\includegraphics[scale=0.6]{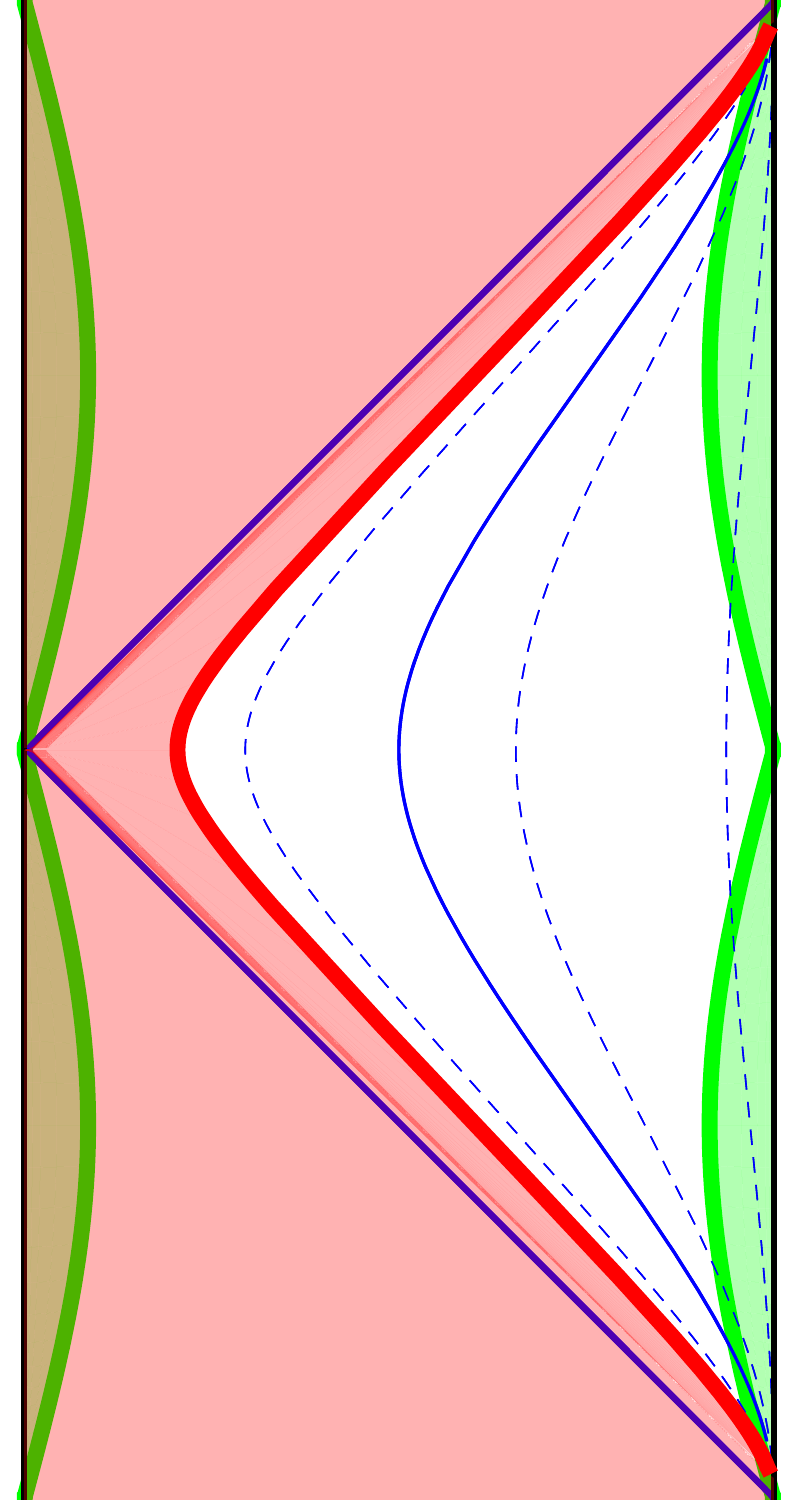}}
\hspace{1cm}
\subfloat[Poincare.\label{palternate}]
{\includegraphics[scale=0.6]{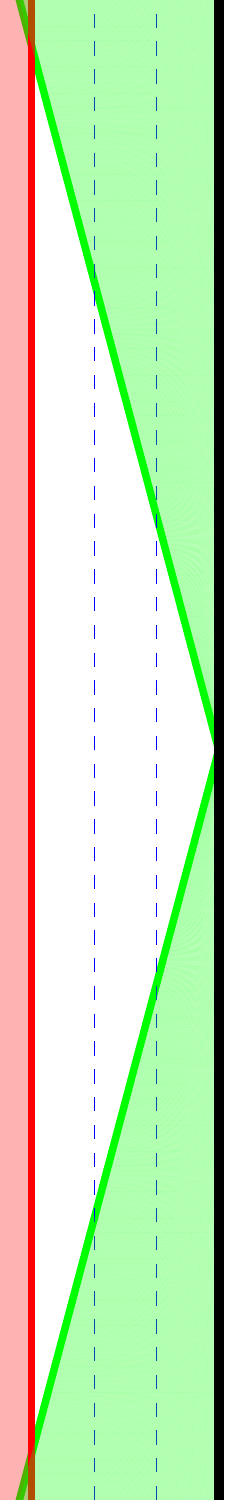}}
\caption{Reborn Universes. Unlike other embeddings displayed in this work, panel c) is not showing a cross section of global AdS, but instead is displaying the same branes as panel b) but using standard Poincare Patch coordinates. \label{bouncinguniverses}}
\end{centering}
\end{figure}

In this spirit, let us study a slight variation of the standard RS setup of figure \ref{standardrs} where we replace the UV brane with two or three dS branes as depicted in figure \ref{bouncinguniverses}.
Recall that in the standard RS setup the IR brane is a negative tension Minkowski brane. We retain this IR brane, but replace the UV brane (which used to be a positive tension Minkowski brane) with either 3 dS branes with $\tau_*=- 3 \pi/2$, $\tau_*=- \pi/2$ and $\tau_*=3 \pi/2$ or with 2 dS branes with $\tau_*=-\pi$ and $\tau_*=0$.
We will refer to the brane world in figure \ref{standardbouncing} as the standard reborn universe, and the one in figure \ref{alternatebouncing} as the alternate reborn universe. It is the latter that will prove the most useful in constructing effective actions.

We already understand how to relate both the Minkowski adapted and the de Sitter adapted coordinates to the global AdS$_5$ coordinates. But for the purposes of understanding the physics of the reborn branes we would prefer yet a different piece of information: What is the radial position of the UV brane as a function of space and time as viewed from the Minkowski brane observer? That is, we need to translate the condition that a brane is living at a constant $r_S$ in the de Sitter adapted coordinates of (\ref{adaptedds}) into a $z(t,x)$ in terms of the standard Poincare Patch coordinates of (\ref{ppatch}). This map is very easy in the case of the alternate scenario of figure \ref{alternatebouncing}. In this alternate scenario the UV and IR branes are never even approximately parallel, so in this scenario one would see strong space and time dependence of Newton's constant throughout. But the geometry is very illuminating and easy. The parameterization (\ref{ppatchembedding}) agrees with the $X_5 \leftrightarrow X_0$ reversed (\ref{flatdsembedding}) if we set
\beq x=x_S, \quad z = e^{-t_S} \, \mathrm{csch} \, r_S, \quad t = e^{-t_S}\, \coth r_S \ee
or in other words
\beq
\label{altsolution}
z = (\mathrm{sech}\, r_S) \, t = \frac{t}{\sqrt{l^2+1}} ,
\eeq
the radial position of the dS brane depends only on the Minkowski time coordinate and grows linearly with it. Here $l$ is the curvature radius on the brane, which is related to $r_S$ via (\ref{hubble}). As expected this yields a phenomenologically unacceptably large time dependence of the brane separation. Given the simple form of this embedding, we chose to also display the alternate embedding in the Poincare Patch coordinates of (\ref{ppatch}) in figure \ref{palternate}. The horizontal direction is $z$ and time $t$ runs vertically. The $x$ and $\omega_2$ directions are suppressed. Since these are the coordinates adapted to the Minkowski brane it just turns into a vertical line in this figure. The simple soltion (\ref{altsolution}) means that the de Sitter branes are diagonal lines.

To describe the reborn branes of figure \ref{standardbouncing} we need to once more take the parameterization (\ref{ppatchembedding}) of the Poincare Patch, but this time equate it with (\ref{globaldsembedding}) as written\footnote{Of course we could also once again compare to (\ref{flatdsembedding}), but this time without reversing $X_5$ and $X_0$, which takes away all the simplifications of this case.}. Unfortunately the result is quite a bit more cumbersome. Let us introduce an angle $\beta$ with $\Omega_4 = \cos \beta$, $\Omega_a = \sin \beta \omega_a$. With this we get
\begin{eqnarray}
z &=& \frac{\text{csch}(r_S) \text{sech}(\tau_S) \sec ^2(\beta) (\coth (r_S) \text{sech}(\tau_S)-\sin (\beta))}{\sec ^2(\beta) \left(\text{csch}^2(r_S) \text{sech}^2(\tau_S)-\tanh ^2(\tau_S)\right)+1}
\nonumber \\
t &=& \frac{2 \cosh (r_S) \tanh (\tau_S) \sec (\beta)-2 \sinh (r_S) \sinh (\tau_S) \tan (\beta)}{2 \sinh (r_S) \cosh (\tau_S) \cos (\beta)-2 \sinh (r_S) \sinh (\tau_S) \tanh (\tau_S) \sec (\beta)+2 \text{csch}(r_S) \text{sech}(\tau_S) \sec (\beta)} \nonumber \\
x & =& \frac{\coth (r_S) \text{sech}(\tau_S) \sec (\beta)-\tan (\beta)}{\sec ^2(\beta) \left(\text{csch}^2(r_S) \text{sech}^2(\tau_S)-\tanh ^2(\tau_S)\right)+1}
\end{eqnarray}
Our task should be to get the UV brane position $z(t,x)$ from this. This seems to be impossible to obtain in closed form. However, expanding around $\tau_S=0$ and $\beta=\pi/2$, the point at which the two branes are almost parallel, we find, dropping terms beyond quadratic order,
\begin{eqnarray}
z & =& e^{-r_S} - \frac{\tau_S^2}{2} e^{-2 r_S} \sinh r_S \, + (\frac{\pi}{2} - \beta)^2 \, \frac{e^{-2r_S}}{2} \sinh r_S  \nonumber \\
t & = &  \tau_S \, e^{-r_S} \, \sinh r_S \,\nonumber \\
x & =& (\frac{\pi}{2} - \beta) \, e^{-r_S} \, \sinh r_S \, .
\end{eqnarray}
To leading (linear) order the dS time and space coordinates $\tau_S$ and $\beta$ are essentially just the Minkowski coordinates $t$ and $x$, up to rescaling, and the $z$ position is constant.
To find the leading time and space dependent correction to $z$ we find at quadratic order that a dS brane at a constant $r_S$ corresponds to $z(t,x)$ of
\beq
\label{bouncingposition}
z = e^{-r_S} + \frac{1}{2 \sinh r_S} (x^2-t^2).
\eeq
Note that the coordinates $x$ and $t$ of the Poincare slicing are not quite the correct space and time coordinate of the IR brane. The induced metric on the IR brane is (recall that we used $x$ as the radial direction of spherical coordinates on Minkowski space)
\beq
ds^2_{IR} = \frac{1}{z_0^2} (- dt^2 + dx^2 + x^2 d \omega_2^2).
\eeq
To get properly normalized coordinates we need to rescale by the IR warp factor, $t_{IR}=t/z_0$, $x_{IR}=x/z_0$. In terms of these variables the position of the planck brane of (\ref{bouncingposition}) expanded to quadratic order near $x=t=0$ reads
\beq
\label{scaledbouncingposition}
z = e^{-r_S} + \frac{z_0^2}{2 \sinh r_S} (x_{IR}^2-t_{IR}^2).
\eeq
The hope here might be to find a solution with small space and time dependence at large $t$ so that we can generate a phenomenologically acceptable solution, even without any additional stabilizing ingredients. However, this is not so simple.
As long as the energy mismatch of the UV brane is small, $\sinh(r_S)$ is exponentially large, but this doesn't help in terms of suppressing time dependence as leading and subleading term have the same exponential suppression. Furthermore, the IR warp factor $z_0$ seems to be working against us as it enhances the spatial variation. As it stands, this scenario also yields unacceptably large spatial and temporal dependence of the brane separation.

Another interesting expansion we can look at in this case is around $\tau \rightarrow -\infty$, the point at which the UV brane bounces off the boundary. In this limit it is straightforward to see that $z$ once more becomes independent of $x$ and is very well approximate by exactly the same linear dependence $z = (t+1)/\sqrt{l^2+1}$ as in eq. (\ref{altsolution}) up to an irrelevant constant shift in $t$. While the full solution strongly depends on $\tau_*$, the solution near the rebirth is universal.

\section{Low energy effective action}

\subsection{Basic Strategy}
In this section we will use the alternate reborn scenario of figure \ref{alternatebouncing} to discuss one more important issue: what is the low energy description of the brane worlds we constructed?
With one dS and one Minkowski brane, what is the metric dominating the low energy world? Is it Minkowski or dS? What is the action from which these solutions follow? Note that an effective action for the low energy dynamics of the coupled metric/radion system had already been worked out previously \cite{Chacko:2013dra}. They find\footnote{We found it convenient to rescale the field $\varp$ of \cite{Chacko:2013dra} by a factor of $\sqrt{12} \kappa$ with $\kappa=2 M_5^3/k=(16 \pi G)^{-1}$ in order to have an overall factor of $\kappa^2$ in front of the action rather than a canonically normalized scalar as in their work.}
\beq
\label{chacko}
S=\frac{1}{16 \pi G} \int d^4x \sqrt{-g} \,\left [ \left (1 - \varp^2 \right ) R - 6
(\partial \varp )^2 - 2 V(\varp)
\right ]
\eeq
with
\beq
\label{chackopot}
V(\varp) =  ( \Lambda_{UV} - \Lambda_{IR}\,  \varp^4)
\eeq
where $\Lambda_{UV/IR}$ are the effective cosmological constants on the brane related to the curvature radius $l$ of the vacuum (A)dS solution on the respective branes via $|\Lambda| = 3/l^2$. Note that the radion couples non-trivially to the Einstein Hilbert term. For the particle physics applications considered in \cite{Chacko:2013dra} this effect was negligible; they were mostly interested in small fluctuations around flat space. Clearly if we want to reproduces the cosmological solutions with genuinely mismatched branes this coupling will be crucial. In fact, the strategy we would like to pursue here is to construct the low energy action of the mismatched branes by engineering an action that allows for the solutions we found. As always when coupling a scalar to gravity an important question is the choice of frame. This frame choice is arbitrary, but we will see that the most useful frame choice is often dictated by whether we want to describe the physics from the point of view of the IR or the UV observer. Reassuringly we will find that in one of our frames the action we construct this way indeed perfectly agrees with the one of \cite{Chacko:2013dra} in (\ref{chacko}).

Before we proceed, a quick word on conventions. The effective action we are after describes a system of a single scalar, the radion, coupled to gravity. We will, however, use several different parameterizations for the scalar. We've already encountered one, $\varp$, which is the radion variable most commonly used in the literature on braneworlds, for example \cite{Chacko:2013dra}. This parameterization however is not very useful when seriously taking into account the couplings of the radion to the Ricci scalar. In general theories of this scalar tensor type it is common, as we'll review below, to use a scalar field variable that couples via $\phi R$ directly to the Ricci scalar. We will introduce a scalar $\phi$ that couples like this in a frame adapted to an observer on the IR brane, and a separate scalar $\tilde{\phi}$ that plays the same role in a frame adapted to the UV brane. The relations between these three scalars are given by (\ref{pptrelation}) and (\ref{radionrelation}). Last but not least we will use the scalar $\tilde{X}$ which is the canonically normalized scalar in the Einstein frame. Its relation to the other variables used for the radion is given by (\ref{einsteinradion}) and (\ref{einsteinradiontwo}).

The point of departure for our study will be the basic observation that in RS brane worlds the 4d Planck scale $M_4$ can be calculated in terms of the 5d Planck scale $M_5$ via
\beq M_4^2 = V M_5^3. \label{mplcond} \eeq
Here $V$ is the warped volume of the internal space. Using the standard Poincare patch coordinates of (\ref{ppatch}) one finds
\beq V = \int dz \, e^{3 A} \sim z_{UV}^{-2} - z_{IR}^{-2}. \label{mpl} \eeq
The explicit solutions for $z$ as a function of space and time in the various coordinate systems we constructed in the previous sections hence can be seen as solutions of a coupled metric/scalar system. We can reverse engineer the action by requiring that these solutions are reproduced.

\subsection{dS-UV brane: simplified scenario with $z_{IR} \rightarrow \infty$}
\label{infzir}

The first scenario we discuss is the alternate reborn universe with a positive tension dS brane in the UV and a negative tension Minkowski IR brane. From the point of view of the known radion potential (\ref{chacko}) we can see that a Minkowski IR brane is particularly simple since it does not involve the $\varp^4$ term in the potential. For the purpose of finding the action the alternate scenario of figure \ref{alternatebouncing} will be easier since we only have to deal with time dependent position $z(t)$.

In order to lay out the basic framework for the effective action, let us first drop the small contribution from the IR brane to $M_4^2 = (16 \pi G)^{-1}$, which is predominantly dependent on $z_{UV}$. Formally this means we first analyse the $z_{IR} \rightarrow \infty$ limit. Since $z_{UV}$ is time dependent we effectively have a time dependent Newton constant $G$, as we emphasized several times before. The Newton constant is set by the vacuum value of a scalar field, the radion, which encodes the separation of the branes. This means the low energy effective action describing these two branes is of the generalized Brans-Dicke or scalar-tensor form\footnote{For a recent review see, for example, \cite{Quiros:2019ktw}}. This immediately brings up the question of frames. We can redefine the metric by powers of the radion to shuffle time dependence from $G$ into the metric. The metric and the action both depend on our choice of frame.

Let us first start with a frame that describes the experience of a Minkowski brane observer. This is the ``Minkowski" frame, which in the Brans Dicke language would correspond to the Jordan frame. In this frame the action is
\beq
S_{BD} = \frac{1}{16 \pi G} \int d^4x \sqrt{-g}\, \left [ \phi R - \frac{w_{BD}(\phi)}{\phi} (\partial \phi)^2
- 2 V(\phi) \right ]. \label{bdaction}
\eeq
Here $\phi$ is the radion field setting Newton's constant. From comparison to eq. (\ref{mpl}) we see that, in the $z_{IR} \rightarrow \infty$ limit we decided to tackle first, we have
\beq \phi = \alpha \, z_{UV}^{-2}. \eeq
We will determine the functions $w_{BD}(\phi)$ and $V(\phi)$ as well as the constant $\alpha$ from the requirement that we reproduce the known solutions from the previous sections. In particular, we know from (\ref{altsolution}) that the alternate reborn scenario requires a solution in which the spacetime is simply Minkowski space and
\beq
\label{phisol}
\phi = \frac{\alpha (l^2+1)}{t^2}  \equiv \frac{a^2}{t^2}.
\eeq
The equations of motion that follow from the action (\ref{bdaction}) are
\begin{eqnarray}
\nonumber
G_{\mu \nu} &=& \frac{w_{BD}}{\phi^2} \left [ \partial_{\mu} \phi \partial_{\nu} \phi - \frac{1}{2} g_{\mu \nu}
(\partial \phi)^2 \right ] - g_{\mu \nu} \frac{V}{\phi} + \frac{1}{\phi} \left (
\nabla_{\mu} \nabla_{\nu} - g_{\mu \nu} \nabla^2 \right ) \phi \\
\nabla^2 \phi &=& \frac{2}{3+2 w_{BD}} (\phi \partial_{\phi} V - 2 V) - \frac{\partial_{\phi} w_{BD}}{3 + 2 w_{BD}} (\partial \phi)^2. \label{bdeom}
\end{eqnarray}
Here $G_{\mu \nu} = R_{\mu \nu} - \frac{1}{2} R g_{\mu \nu}$ is the Einstein tensor (and not the 5d metric, which sometimes is also denoted as such). It is easy to see that $g_{\mu \nu} = \eta_{\mu \nu}$ (and hence $G_{\mu \nu}=0$) together with $\phi$ given by (\ref{phisol}) is a solution if we chose
\beq w_{BD} = - \frac{3}{2}, \quad V(\phi) = \frac{3 \phi^2}{a^2} . \label{parameters} \eeq
That is, the action is already fixed up to an overall constant $a$ by requiring that we reproduce the known solution.

We now change frame to that of a de Sitter observer. To do so, we perform a field redefinition
\beq \label{cof} g_{\mu \nu} \rightarrow  \Omega^{-2} g_{\mu \nu} \eeq
where we chose
\beq \Omega^2 = \phi. \eeq
The physics of the system can not change since this was solely a field redefinition, but it changes what we mean by the metric. In this new ``de Sitter frame" the metric reads
\beq
\label{bdds}
ds^2 = \frac{a^2}{\, t^2} (-dt^2 + d \vec{x}^2).
\eeq
This is indeed dS space with curvature radius $a$. Since we want $a=l$ to agree with what we see on the dS brane we can fix our last parameter
\beq a = l, \quad \Rightarrow \quad \alpha = \frac{l^2}{l^2+1}.\eeq
The action in the de Sitter frame then reads
\begin{eqnarray}
S_{BD}^{dS} &=& \frac{1}{16 \pi G} \int d^4x \sqrt{-g}\, \left [ R - \frac{w_{BD} + \frac{3}{2}}{\phi^2} (\partial \phi)^2
- 2 \frac{V(\phi)}{\phi^2} \right ] = \nonumber \\
&=& \frac{1}{16 \pi G} \int d^4x \sqrt{-g}\, \left [ R
- \frac{6}{l^2}  \right ] \label{nokinds}
\end{eqnarray}
where in the last step we used our identifications of $w_{BD}$ and $V$ from (\ref{parameters}).

$G$ is constant in this new frame in which the potential is just a positive cosmological constant and the radion has completely dropped out of the action. Clearly this action allows for a dS solution with curvature radius $l$.
We apparently lost all time dependence in the solution. This is of course not the case. The radion is still given by the same solution (\ref{phisol}). However, it no longer couples to the gravitational sector. This is due to the fact that we are working with $z_{IR} \rightarrow \infty$ and so the energy of the Minkowski brane is warped down to zero. Matter living on the Minkowski brane would see this time-dependence through the $\phi(t)$ appearing in its coupling constants\footnote{Matter on the IR brane couples to the Minkowski frame $g_{\mu \nu}$ without any powers of $\phi$, so that after the field redefinition into the dS frame it now couples to the dS frame metric with additional powers of $\phi$, inheriting strong explicit time dependence.}.

So already in this simple scenario we see that we can describe one and the same physics in two very different ways. We either can describe our branes as a Minkowski space with a strongly time dependent Newton's constant (due to the motion of the UV brane) or as a dS space with a constant $G$ but strong time dependence in the couplings of any IR localized matter sector. In order to account for the gravitational backreaction of the latter we need to work at a finite $z_{IR}$. We will turn to this next.

\subsection{Finite $z_{IR}$}
\label{unlimitedeinstein}

To study the full effective action, we redo the analysis of the previous subsection in the presence of a finite $z_{IR}$. In RS scenarios $z_{IR}$ is usually thought of as an exponentially large IR warp factor that leads to a large hierarchy of scales between the Planck physics on the UV brane and TeV scale physics on the IR brane. Our analysis will not rely on $z_{IR}$ as being large; the effective action we derive is valid for all $z_{IR}$. But the language we use to interpret the results will assume a large $z_{IR}$. The results of the previous subsection follow in the $z_{IR} \rightarrow \infty$ limit.

The upshot of this analysis will be that there are actually {\it three} different frames that are useful to describe the physics of the reborn scenarios:
\begin{itemize}
\item
the Minkowski frame from above, where the spacetime is Minkowski space and we have a strongly time dependent $G$ from the relative motion of the UV dS brane
\item
the de Sitter frame from above in which the spacetime is dS and we retain a weak (that is $z_{IR}$ suppressed) time variation of $G$ from the relative motion of the IR Minkowski brane
\item the genuine Einstein frame in which $G$ is constant, but the four-dimensional spacetime metric is neither Minkowski nor dS but some ``average" of the two though deviations from dS are $z_{IR}$ suppressed.
\end{itemize}

As in the previous subsection, let us start with the Minkowski frame. Once again we want to write a general scalar tensor action of the form (\ref{bdaction}), but this time we require a Minkowski space solution when the solution of $\phi$ takes the form
\beq
\label{phisoltwo}
\phi = \alpha (z_{UV}^{-2} - z_{IR}^{-2}) =\frac{l^2}{t^2} - c_0
\eeq
where we defined
\beq
c_0 = \frac{l^2}{l^2+1} z_{IR}^{-2}.
\eeq
The $z_{IR} \rightarrow \infty$ limit corresponds to $c_0=0$. It is straightforward to confirm that this metric and scalar field are solutions to the BD equations of motion (\ref{bdeom}) only if we choose
\beq
w_{BD} = - \frac{3}{2} \frac{\phi}{ \phi + c_0}, \quad V(\phi) = \frac{3 (\phi+c_0)^2}{l^2} . \label{parameterstwo}
\eeq
This has the right $c_0=0$ limit in which it reduces to (\ref{parameters}). Note that, despite appearances, $c_0$ should be thought of as an integration constant rather than a parameter in the action. But we found that if we study fluctuations around the $z_{IR}$ solution, the radion ended up without a kinetic term in the de Sitter frame. Here we effectively introduced a shifted radion that describes fluctuations around the IR brane whose location at $t=0$ is parameterized by $c_0$. The integration constant $c_0$ now appears in the potential of the shifted radion. As we will see in the end if we go to the genuine Einstein frame any $c_0$ dependence will disappear from the action.

Next let us turn to the dS frame. For the metric to be dS we need to do the same conformal transformation that gave us the metric (\ref{bdds}). That is we need to choose
\beq
\label{goodomega}
\Omega^2 = \frac{l^2}{t^2} = \phi + c_0.
\eeq
Note that this means that the dS frame is no longer the Einstein frame. The 4d $G$ is dominated by the UV brane position, which in the dS frame this is effectively constant. Nonetheless, according to (\ref{mplcond}) there is a mild dependence on $z_{IR}$ as well so the motion of the IR brane imprints a small variation of $G$ even in the dS frame. To get rid of all time dependence of $G$ we would need to continue using $\Omega^2=\phi$. We will do this later.
The action in the dS frame reads
\begin{eqnarray}
S_{BD}^{dS} &=& \frac{1}{16 \pi G} \int d^4x \sqrt{-g}\, \left [\frac{\phi}{\phi+c_0} R - \frac{3 c_0}{2}
\frac{ (\partial \phi)^2}{(\phi + c_0)^3}
- \frac{6}{l^2} \right ] = \nonumber \\
&=& \frac{1}{16 \pi G} \int d^4x \sqrt{-g}\, \left [ \tilde{\phi} R+
 \frac{3}{2} \frac{\tilde{\phi}}{\tilde{\phi}-1}\frac{(\partial \tilde{\phi})^2}{\tilde{\phi}}  -\frac{6}{l^2}  \right ]
\label{fulldsframeaction}
\end{eqnarray}
where we introduced
\beq \tilde{\phi}=\phi/(\phi+c_0)
\label{pptrelation}
\eeq
to bring the action back into the standard scalar-tensor form of (\ref{bdaction}). It is in this form that we can recognize our action to be identical to the one of \cite{Chacko:2013dra} as written in (\ref{chacko}) above using the substitution
\beq \tilde{\phi} = 1 - \varp^2. \label{radionrelation} \eeq

It is reassuring that we recover this known result from our orthogonal approach. Our analysis also serves to put the action (\ref{chacko}) of \cite{Chacko:2013dra} in the right perspective. We should think of this action as written in a frame {\it appropriate for an observer on the UV brane} with the field $\varp$ setting the {\it position of the infrared brane}, $\varp \sim z_{IR}^{-1}$. This perspective will be important when deriving the radion mass. Most importantly, the field configuration $\varp=0$ corresponds to an IR brane at the horizon, $z_{IR} \rightarrow \infty$. Static 2-brane systems with matched tensions are solved with constant, finite $\varp$.

Note that in the $c_0 \rightarrow 0$ limit $\tilde{\phi}$ remains close to 1 for almost the full range of $\phi$. That is as expected since in the dS frame the radion dynamics almost decouples. There is however a small time dependence of the Planck scale owing to the motion of the IR brane. Moreover, as before, if additional matter is localized on the IR brane, it sees strong time dependence from $\phi$. It is straightforward to confirm\footnote{Note that the 2nd derivative in dS is non-trivial due to the Christoffel symbols, \beq
\nabla_t \nabla t f(t) = \frac{\dot{f}}{f} + \ddot{f}, \quad
\nabla_i \nabla_j f(t) = \delta_{ij} \frac{\dot{f}}{f}. \eeq} that the equations of motion following from the action (\ref{fulldsframeaction}), which once again take the form of (\ref{bdeom}), indeed are solved by a metric that is dS with curvature radius $l$ and a scalar given by
\beq
\tilde{\phi} = \frac{\phi}{\phi+c_0} = 1 - c_0 \frac{t^2}{l^2}.
\eeq

Last but not least we can go to the actual Einstein frame. That is, we start again with (\ref{bdaction}) and this time we indeed choose
\beq
\Omega^2 = \phi
\eeq
instead of (\ref{goodomega}) above.
In the Einstein frame the action becomes
\begin{eqnarray} \nonumber
S_{BD}^{E} &=& \frac{1}{16 \pi G} \int d^4x \sqrt{-g}\, \left [R - \frac{3 c_0}{2 \phi^2}
\frac{ (\partial \phi)^2}{(\phi + c_0)} - \frac{6}{l^2} \frac{(\phi+c_0)^2}{\phi^2} \right ]  \\
&=&  \frac{1}{16 \pi G} \int d^4x \sqrt{-g}\, \left [R -
 (\partial \tilde{X})^2 - \frac{6}{l^2} \cosh \left ( \frac{\tilde{X}}{\sqrt{6}} \right )^4 \right ] \label{limitedeinstein}
\end{eqnarray}
where in the second step we introduced
\beq
\tilde{X} = \sqrt{\frac{3}{2}} \log \left ( \frac{\sqrt{\phi +c_0} + \sqrt{c_0}}{\sqrt{\phi+c_0} - \sqrt{c_0}} \right ) \label{einsteinradion}
\eeq
in order to work with a canonically normalized scalar in the Einstein frame. Note that the $c_0$ dependence has scaled out from the action. It is useful to note that the relation (\ref{einsteinradion}) implies a very simple relation between $\tilde{X}$ and $\varp$:
\beq \varp = \tanh \left ( \frac{ \tilde{X}}{\sqrt{6}} \right ) \label{einsteinradiontwo} .\eeq

In the Einstein frame the metric is neither dS nor Minkowski, but instead is given by
\beq \label{almostdsmetric} ds^2 = \left ( \frac{l^2}{t^2}- c_0 \right ) (-dt^2 + d \vec{x}^2) .\eeq
As expected from the previous subsection, this reduces to the dS metric in the $c_0=0$ limit, but in general this metric deviates from dS at late times. It is reassuring to confirm that the standard Einstein equations with a $\cosh^4$ potential for the scalar $\tilde{X}$ are indeed solved by the metric (\ref{almostdsmetric}) together with the scalar profile
\beq
\tilde{X}=\sqrt{\frac{3}{2}} \log \left ( \frac{l + \sqrt{c_0} t}{l - \sqrt{c_0} t} \right )
\eeq
which we inherit from the solution for $\phi$ in (\ref{phisoltwo}).

There are additional consistency checks we can perform to confirm that this effective action is indeed correct. For example, one can generalize the scenario to dimensions other than 4d branes in 5d AdS. While we haven't checked all the order 1 factors, it is assuring that the obvious scaling with dimensions works out fine. Working with a parent AdS$_{d+1}$ the warped volume of equation (\ref{mpl}) now will be dominated by $z_{UV}^{2-d}$ and correspondingly the solution for $\phi+c_0$ will scale as $(\phi+c_0) \sim t^{2-d}$. The conformal transformation to the dS frame picks up a factor of $\Omega^{-d+2} = (\phi+c_0)^{\frac{-d+2}{2}} \sim t^2$ and we find that indeed in any $d$ a Minkowski metric in the Minkowski frame turns into a dS metric in the dS frame.

\subsection{Detuned IR branes}

One of the interesting aspects of the action (\ref{chacko}) that we haven't touched upon yet is the presence of a $\varp^4$ potential for detuned IR branes. In this subsection we would like to show that when we detune the IR brane this term is indeed required in the action in order to reproduce our non-trivial brane configurations.

In order to analyze this scenario we will reverse the role of the UV and IR branes. That is, we use the Minkowski adapted coordinates of (\ref{ppatch}) in which the position of the dS brane with $\tau_*=0$ is still given by (\ref{altsolution}). But this time the dS brane is taken to be the IR brane
\beq z_{IR}= \frac{t}{\sqrt{l^2+1}} \eeq
and we place an additional Minkowski brane at a constant $z_{UV}<z_{IR}$. That is, this time our radion solution is given by
\beq \label{newsol} \phi =C_0 - \frac{l^2}{t^2} \eeq
instead of (\ref{phisoltwo}). Here $C_0 \sim z_{UV}^{-2}$.
Using a frame adapted to the UV Minkowski brane this is now supposed to be a solution to the equations of motion following from the action that now includes an extra $\varp^4$ term, which in terms of the $\tilde{\phi}$ variable reads
\begin{eqnarray}
S_{BD}^{UV}
&=& \frac{1}{16 \pi G} \int d^4x \sqrt{-g}\, \left [ \tilde{\phi} R+
 \frac{3}{2} \frac{\tilde{\phi}}{\tilde{\phi}-1}\frac{(\partial \tilde{\phi})^2}{\tilde{\phi}}  + \frac{6 C_0}{l^2}  (1-\tilde{\phi})^2 \right ] .
\label{fulluvframeaction}
\end{eqnarray}
This is to be compared with the de Sitter frame action from the reborn universes (\ref{fulldsframeaction}). The only difference is that the positive constant potential from the detuned UV brane has been replaced by a negative $\varp^4 = (1-\tilde{\phi})^2$ potential from the detuned IR brane. The extra factor of $C_0 \sim z_{UV}^{-2}$ in the potential appears since we here write the IR cosmological constant in terms of the curvature radius $l$ of the UV-brane dS space. The time dependence of the Newton constant implied by (\ref{newsol}) as seen from the UV brane reads
\beq \tilde{\phi} = 1 - \frac{l^2}{C_0 t^2} .\eeq
It is straightforward to confirm that this scalar profile together with a Minkowski metric is, in fact, a solution to the equations of motion following from (\ref{fulluvframeaction}). This is a highly non-trivial check of the framework laid out here and in \cite{Chacko:2013dra}.

\subsection{Radion Mass}
\label{radionmass}

Another very interesting application of the action (\ref{chacko}) is to return to the case of matched branes -- that is two de Sitter or two Anti-de Sitter branes. Let us denote the corresponding curvature radii $l_{IR}$ and $l_{UV}$. In this case we know that solutions should exist with constant $\varp$. The radion mass for two matched branes had been analyzed in \cite{Chacko:2001em} with the result that
\beq \label{fox} m^2 = - \frac{4}{l_{IR}^2} \, \mbox{ for dS}, \quad \quad m^2 = + \frac{4}{l_{IR}^2}
\, \mbox{ for AdS}. \eeq
At first sight this appears puzzling since the $-R \varp^2$ in the action (\ref{chacko}) indicates that a positive scalar curvature gives the radion a {\it positive} mass squared and vice versa for a negative curvature, which appears to be direct contradiction with the results quoted in (\ref{fox}). The resolution of this puzzle is that $\varp=0$ is the wrong extremum of $V$. Recall that for us to have two dS or two AdS branes the IR brane is necessarily detuned and hence we have a non-vanishing $\varp^4$ term. In the dS case (with $R=+12/l_{UV}^2$) the full radion dependent part of the potential reads:
\beq
\label{radionpotential}
 V = 3 \left ( 2 \frac{\varp^2}{l_{UV}^2} - \frac{\varp^4}{l_{IR}^2} \right ) .\eeq
and $V \rightarrow -V$ for the AdS case.

In the dS case, $\varp=0$ is a metastable minimum. But this is not the point describing the standard two-brane scenario; instead it describes the case where one brane fell into the $z \rightarrow \infty$ horizon. The standard case places the IR brane at the unstable maximum. Indeed we can verify that\footnote{The prefactor of 6 in the relation between $m^2$ and $V''$ comes from the unconventional normalization of the kinetic term in (\ref{chacko}).}
\beq V'((\varp)_*)=0 \, \mbox{ for } \, (\varp)_*^2=  \frac{l_{IR}^2}{l_{UV}^2}, \quad \quad
m^2 = \frac{1}{6} V''\left ( (\varp)_* \right ) =- \frac{4}{l_{UV}^2} . \label{value} \eeq
For two AdS branes we similarly have an unstable maximum at $\varp=0$. Here the true two-brane solution corresponds to the stable minima at $(\varp)_*$ with $m^2=+4 l_{UV}^{-2}$. This result at the stable minimum now looks almost consistent with (\ref{fox}) except for the $l_{UV} \leftrightarrow l_{IR}$ exchange. To understand this rescaling, one should note that the work of \cite{Chacko:2001em}, in contrast to \cite{Chacko:2013dra}, was done in a frame adapted to the IR brane. The rescaling of the metric from the UV-frame to the IR-frame introduces exactly a factor of $l_{UV}^2 l_{IR}^{-2}$. Once again, it is very reassuring to see known results pop out of the radion action in a highly non-trivial fashion\footnote{Note that when we want to calculate the effective cosmological constant the potential as written is not helpful. The quadratic term is the curvature coupling, so it is not part of the potential energy that backreacts on the metric. Furthermore, we also need to include the constant $3 l_{UV}^{-2}$ contribution. The full potential at the unstable maximum hence is $3 l_{UV}^{-2}(1 - l_{IR}^2 l_{UV}^{-2})$, which is exactly what is needed to support a dS space with curvature radius $l_{UV}$ with the reduced $(1-\varp^2) R$ Einstein-Hilbert term.}.

\subsection{Generalizations}

Having demonstrated the power of the radion effective action in describing the full non-linear structure of mismatched brane setups, we can explore other (and in some cases more physical) setups. One important generalization is to replace the bulk AdS$_5$ background with an AdS$_5$ Schwarzschild black hole. This corresponds to giving a finite temperature to the dual CFT. Here we want to demonstrate that the radion effective action still suffices to describe the brane geometry.

The full 5d brane geometries are known. The analysis for a single brane has been done in all generality a long time ago in \cite{Kraus:1999it}. The 5d geometry is given by
\beq
ds^2 = - f(R) dt^2 + \frac{dR^2}{f(R)} + R^2 d\Sigma_k^2
\eeq
where $d\Sigma_k^2$ describes the metric on the three dimensional unit sphere, flat space or hyperboloid respectively depending on whether $k=+1$, $k=0$ or $k=-1$. The blackening function $F$ is given by
\beq F(R) = k + R^2 - \frac{\mu}{R^2}. \eeq

For $\mu=0$ these are just yet different coordinate systems for AdS$_5$. Non-zero $\mu$ introduces a bulk black hole (and hence non-zero temperature). From the holographically dual perspective $\mu$ is proportional to the energy density. Assuming the branes respect the spatial symmetry of the background, their embedding is given by a function $R(t)$ leading to an induced metric
\beq
ds^2 = - d \tilde{\tau}^2 + R^2 d \Sigma_k^2
\eeq
where \cite{Kraus:1999it} reparameterized time to make sure that the metric takes an FRW form. $R(\tilde{\tau})$ obeys the standard Friedmann equation with the brane tension mismatch providing a cosmological constant and the bulk black hole a radiation component. In fact, for our purposes here we will need to use yet another time coordinate, so that the metric reads
\beq
\label{conformaltrafo}
ds^2 = R^2 (- dt_{FRW}^2 + R^2 d \Sigma_k^2)
\eeq
with
\beq
\label{timetrafo}
R \, dt_{FRW} = d \tilde{\tau}.
\eeq
The equations of motion for $R(t_{FRW})$ are less familiar, and in many interesting cases there is no simple closed form solution to the coordinate transformation of (\ref{timetrafo}). What is important though about the metric (\ref{conformaltrafo}) is that the induced metric on separate branes with embeddings $R_{UV}(t_{FRW})$ and $R_{IR}(t_{FRW})$ are simply related to each other by a conformal transformation.  In particular, the $\tilde{\tau}$ variable is different for the UV and IR brane, whereas both branes use the same conformal time $t_{FRW}$. The form (\ref{timetrafo}) makes it easy to read off the radion profile, given a known solution of the Einstein equations. In fact, we claim that the correct identification is
\beq
\varp = \frac{R_{IR}(t_{FRW})}{R_{UV}(t_{FRW})}.
\eeq
This formula is the precise implementation of our previous intuitive identification of the radion from the warped volume of the internal space. The correct, general rule is that the radion corresponds to the conformal factor relating the worldvolume metrics of the two branes.

It is straightforward to check that this prescription indeed allows us to rediscover our old solution and construct many new solutions of the radion effective action (\ref{chacko}). For example
\begin{itemize}
\item For two Minkowski branes with $k=0$ $R_{IR}$ and $R_{UV}$ are constants and hence so is the radion.
\item For a dS UV brane we have $R_{UV}= l e^{\tilde{\tau}/l}$ for $k=0$ and $R_{UV} = l \sinh(\tilde{\tau}/l)$ for  $k=-1$. The corresponding FRW metrics solve the coupled Einstein-radion equations of motion for $\varp = l_{IR}/l$ if the IR brane is dS as well, $\varp=a_0 e^{- \tilde{\tau}/l}$ if the IR brane is Minkowski and $k=0$, $\varp = a_0 \cosh^{-2}(\tilde{\tau}/(2l))$ if the IR brane is Minkowski and $k=-1$, and $\varp=l_{IR}/(l \cosh(\tilde{\tau}/l))$ when the IR brane is AdS and $k=-1$. $a_0$ in these expressions is an integration constant associated with the IR brane position. Here we used the ratio of $R_{IR}$ and $R_{UV}$ in terms of $t_{FRW}$ to calculate $\varp$, but then translated the final answer back into a function of the $\tilde{\tau}$ coordinate of the UV brane. While the coordinate system used here is very different from what we worked with so far, these are just our old mismatched solutions in yet another form.

\item If we turn on a finite temperature, $\mu \neq 0$, and study two critical tension branes we find a solution with $R_{UV} = \sqrt{b_0^2 + 2 \mu^2 t}$ and constant $\varp$. $b_0$ is again an integration constant, this time associated with the UV brane position.
\end{itemize}
In the last case we need to include the contribution of a diagonal traceless stress tensor with spatial entry $p = \mu R_{UV}^{-4}$ in the Einstein equations as appropriate for radiation. What is important here is that we did not need to include any $\mu$ dependent corrections to the potential of $\varp$. The effects of finite temperature  give rise only to the extra radiation and do not modify the radion potential in this case of two critical branes (where the zero temperature potential vanishes identically).

In fact we can understand the above result by considering the limits of large and small temperature. Any temperature depedence can correct the potential only through terms of the form $T l_{IR/UV}$ for one or the other cosmological constants or on the inverse of this function.   For flat tuned branes there would be no such terms. With detuned branes, we can rule out such dependence as well. The latter possibility which wouldn't have sensible $T=0$ behavior. The first terms don't have sensible flat space behavior. It would be nice to see this explicitly in our formalism. Unfortunately in this case the change of coordinates from $\tilde{\tau}$ to $t_{FRW}$ involves inverses of hypergeometric functions and it is difficult to extract information about the potential from the known solutions.

\subsection{Implications}
Perhaps the most important lesson to be learned from our construction is that a fact that appears obvious from a 5d perspective -- that observers on the IR brane and UV brane would describe one and the same brane setup very differently -- is nicely accounted for in terms of the 4d effective action. The different 5d perspectives correspond to different choices of frame in the 4d description. The analysis also drives home our previous point that the time dependence signals trouble for the scenario of \cite{Banerjee:2018qey,Banerjee:2019fzz}. Unlike our reborn universes, the authors of these papers have a UV Minkowski brane and a dS IR brane. So the choices are either to work in a frame in which one has the desired dS metric, but at the cost of a very strong time dependence of $G$ due to the moving UV brane, or to work in a frame with a mild time dependence of $G$, but a metric that is simply Minkowski space.

Another important point is that to eliminate time dependence today, a realistic cosmology based on mismatched branes almost certainly needs implementation of a Goldberger Wise (GW)-type mechanism \cite{Goldberger:1999uk} to stabilize the branes. This is clearly an important next step  and we will describe some attempts in this direction in the next section. The radion effective action will be instrumental in doing so.

To genearate more conventional inflationary scenarios, the universe would start away from the boundary. Notably however, the solutions we have  presented introduce another, more speculative  possiblity, namely that  the reborn scenarios of figure \ref{bouncinguniverses} could address some cosmological puzzles by crucially making use of the time-dependent Newton constant. As we laid out, the full time-dependent solution can be completely captured by a low energy effective action for the radion-metric system. As such, one may expect that one can always go to the Einstein frame, with simply the cosmology of a single scalar field coupled to gravity.

This procedure fails however near the point at which the de Sitter brane hits the boundary. For realistic applications, we can always think of $z_{IR}^{-1}$ as an exponentially small quantity and hence the frame adapted to the UV de Sitter brane in which the action (\ref{chacko}) is valid is a very close approximation to the Einstein frame. In this frame the metric is dS. The rebirth happens at $t_S = + \infty$ from the point of the dying dS universe and at $t_S=-\infty$ from the point of the newly created dS brane. The metric at this point from either perspective is infinite.

So in this frame the physics of the rebirth we wish to describe is problematic, not just because it happens at infinite time (and so we really can't describe the transitions to what happens after $t_S=\infty$), but also because the metric in this region is singular. Maybe more to the point, the geometric interpretation in terms of branes moving in AdS$_5$ makes it clear that at the moment of rebirth when $\varp=1$, we need to supplement the action with extra boundary conditions. This is interesting in that it is something that is not obvious in the four-dimensional description. In the 5d geometric setup, we need to specify what happens to the dS brane when it hits the boundary. It is of interest to note that the universe bouncing back from the boundary is a contracting universe. Whether this freedom allows for interesting new cosmological solutions is a question we leave for future investigation.

\section{The Einstein Frame Action and Inflation.}

As we've seen in all our solvable examples, mismatched branes contain some of the key ingredients for cosmological models. But the strong time dependence of all our solutions means that any realistic model needs  eventually to be stabilized, presumably via the
Goldberger Wise (GW) mechanism \cite{Goldberger:1999uk} or a variant thereof. GW stabilizes the branes by introducing an extra scalar field, the Goldberger Wise field $\chi$, with  non-trivial boundary conditions on both branes. The corresponding gradient energy depends on the brane separation and hence gives an extra contribution to the radion potential.

One thing to note is that we are assuming zero temperature throughout our analysis. We also assume the presence of both branes in our initial conditions. In both these critical respects our analysis differs from Ref. \cite{Creminelli:2001th}.  Note that from the point of view of the CFT our initial condition is not an equilibrium state and the presence of the second brane in the UV seems very unnatural. Yet from the perspective of the five-dimensional theory there is no issue.

A full analysis of a complete cosmological scenario with mismatched branes coupled to a GW field is beyond the scope of this current work, even though we hope  to return to it in the near future. But one important last step we want to take in this work is to recast our final answer for the effective action in the Einstein frame. This way the mismatched brane scenario is framed in the language suited to most cosmological discussions: a single scalar coupled to standard Einstein gravity \footnote{This is true with the caveat that any cosmological scenario where the branes hit the boundary has to be done more carefully with additional parameters and boundary conditions.}.

 To translate our effective action to the Einstein frame we need to use the mapping (\ref{einsteinradiontwo})
\beq \varp = \tanh \left ( \frac{ \tilde{X}}{\sqrt{6}} \right ) \equiv \tanh X. \eeq
The Einstein frame potential for the radion is
\beq V = \frac{3 }{l_{UV}^2}  \cosh^4 X - \frac{3 }{l_{IR}^2} \, \sinh^4 X .\eeq
Here the sign choices are for supercritical branes. For subcritical branes we would send $l \rightarrow il$ for the corresponding brane.
Note that what appeared as a constant plus a quartic in (\ref{chackopot}) turned into exponentially growing terms. This is due to the fact that as the branes approach each other (corresponding to $\varp \rightarrow 1$ or $X \rightarrow \infty)$, the effective Newton constant grows and so all energies backreact more strongly. In the Einstein frame this explicitly appears in the potential.

\begin{figure}
\begin{centering}
{\includegraphics[scale=0.6]{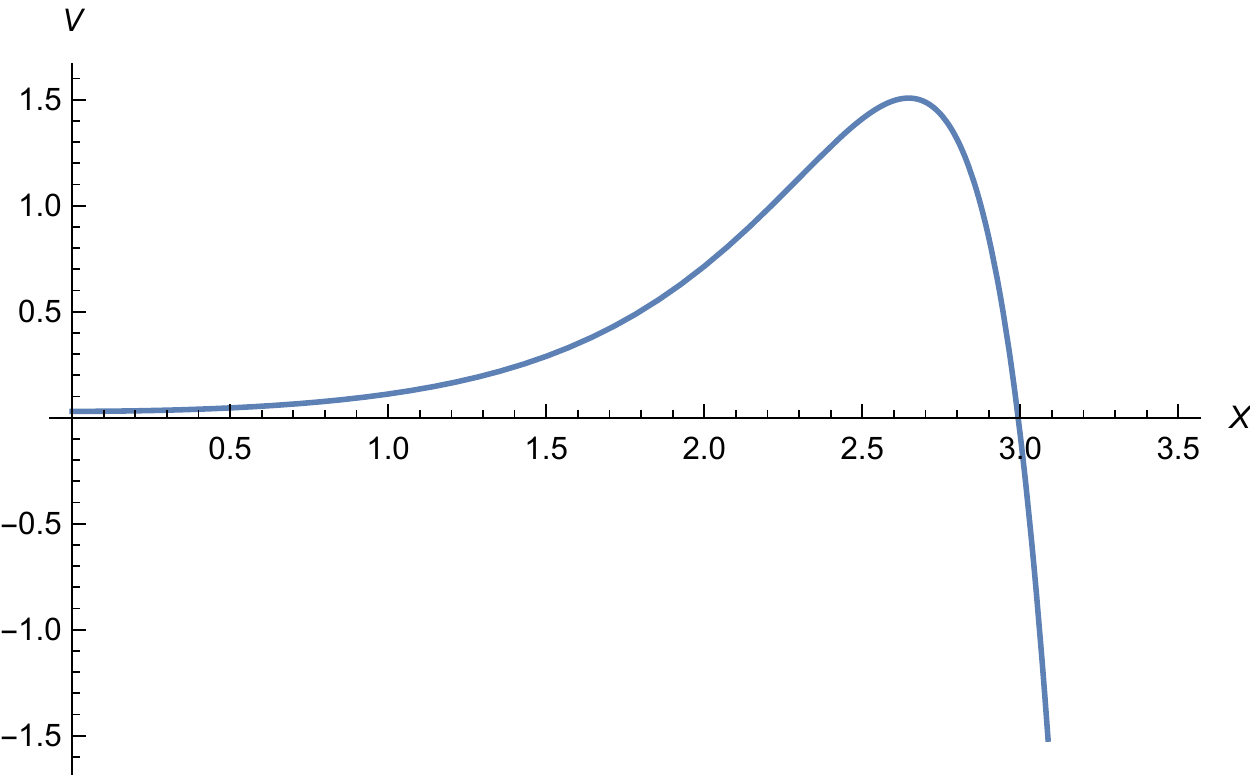} }
\caption{Einstein frame radion potentials for $l_{UV}=10$ and $l_{IR}=9.9$.
\label{einsteinpotentials}}
\end{centering}
\end{figure}

The most interesting scenario that can serve as a starting point for a discussion of standard inflation has both branes are supercritical. As discussed in section \ref{radionmass}, the potential generated by the cosmological terms on the branes has a local minimum at $X=0$ and a maximum at a finite value of $X$ given by (\ref{value}). As $l_{IR} \rightarrow l_{UV}$ (with $l_{UV} > l_{IR}$ being implied by the ordering of the branes) the maximum gets pushed to larger values of $X$ and to higher values of the potential. Beyond the maximum the potential  is unbounded from below due to the negative contribution of the IR brane potential. In figure \ref{einsteinpotentials} we plot the potential for $l_{UV}=10$ and $l_{IR}=9.9$

This potential is tantalizing from the perspective of inflationary dynamics, since in principle the radion can start with high energy and roll down to a stable minimum with small cosmological constant.   Furthermore there could in principle be a natural explanation for the higher energy initially since Newton's constant initially can be much smaller if the cosmological constants on the two branes are close in value. This can be consistent with perturbativity if the AdS scale of the five-dimensional theory is much smaller than the cutoff scale.   One interesting aspect of this scenario is that it is natural for the two branes to have similar parameters if they are created at the same time.

We can study this scenario with  the standard tools. For example, the  scalar could start  just to the left of the peak and subsequently  roll towards its minimum. As it stands, the potential has several shortcomings. For one, the minimum at zero is troublesome; as the IR brane keeps falling towards the horizon matter couplings on the brane will experience strong time dependence. A GW-like scalar is needed to stabilize the brane at a small but finite value. Furthermore, this potential does not allow for standard slow roll. While $V'/V$ can be arbitrarily small by starting the scalar near the top of the potential, $V''/V$ is not small near the peak due to the exponential nature of the potentials.

These are not necessarily unsurmountable obstacles. The rolling of the field could in principle be slowed through additional friction terms since the radion couples to many other fields. As alluded to above, the field can in principle stop for small but nonzero $\varp$ in the presence of a GW field.  This also raises issues with the formulation of the GW potential, which pins down boundary conditions in the UV and the IR. This obscures any correspondence to the dual theory, in which boundary conditions are specified solely in the UV.  The introduction of a GW field seems to destroy the unstable maximum in the simplest models. Yet the GW mechanism should have significant effects on in the IR. Furthermore, higher order terms in the GW potential can significantly affect the GW potential shape--something we don't usually consider when just focusing on a GW minimum in the IR.

Notice also that in any model based on GW, the constraint that if the UV energy doesn't decay, the UV energy would (almost) cancel the energy of the GW field requires that the UV energy density is no greater than the IR scale, which means that the reheat temperature after inflation would also be expected to be of order the IR scale. This may or may not lead to cosmological issues of the sort considered in Ref. \cite{Creminelli:2001th}.

\section{Conclusion} In this paper we have considered an interesting new class of time-dependent brane solutions in AdS space. We have shown explicit solutions and how they are realized in an effective lower-dimensional theory. We have also highlighted some interesting unanticipated features of the low-energy theory--in particular the decoupling of gravity, the necessity to specify brane creation time, and boundary conditions for branes that hit the boundary in a finite time from the perspective of an IR brane observer. We have also pointed out some interesting potential cosmological applications but leave the completion and study of  a successful model to the future.

\section*{Acknowledgments}
We'd like to thank Prateek Agrawal, Hao Geng, and Rashmish Mishra for useful discussions.
The work of AK was supported, in part, by the U.S.~Department of Energy under Grant No.~DE-SC0011637 and by a grant from the Simons Foundation (Grant 651440, AK). The work of LR is supported by NSF grants PHY-1620806 and PHY-1915071, the Chau Foundation HS Chau postdoc support award, the Kavli Foundation grant ``Kavli Dream Team", and the Moore Foundation Award 8342.

\begin{appendix}
\section{Hyperbolic Slicing}
\label{hyperbolic}

Clearly no static solutions exist in the case of mismatched tensions, so we will be forced to look at solutions with time dependent branes. In general this isn't sufficient: for generic mismatched branes the position will be both time {\it and} space dependent. In this appendix we want to show that there exists a special coordinate system in which all three, the Minkowski brane, the AdS brane as well as the dS branes with $\tau_* = - \pi/2$ can be realized as branes with a time but not space dependent position.

If the position of the brane is time dependent, the induced metric on the brane will be an FRW universe whose scale factor is induced by the time dependent brane position. In order to accommodate all three types of branes in a single coordinate system we need to use an hyperbolical spatial slice for the FRW universe since this is the only slicing that allows us to write all 3 maximally symmetric spaces in FRW form. AdS is the troublemaker here: with a negative cosmological constant as the only matter, only a hyperbolic sliced FRW universe is possible. Correspondingly we write the bulk AdS$_5$ metric as
\begin{equation}
\label{AdSmetric}
ds^2 = \frac{1}{z^2} (-dt_h^2 + t_h^2 dH^2 + dz^2).
\end{equation}
Here $dH^2$ denotes the metric on the unit 3d hyperbolic space. The metric in (\ref{AdSmetric}) is just the standard Poincare slicing with Minkowski space on each $z=const.$ slice. But the Minkowski space is written in a peculiar form, basically Wick rotated spherical coordinates in which the radial direction is timelike. This way Minkowski space appears as a particular FRW universe.

A Minkowski RS brane in this geometry still simply sits at a constant
\begin{equation}
z_M(t_h)=z_0.
\end{equation}
Our claim is that both the AdS and the dS branes can be written in these coordinates as well by allowing $z$ to be time dependent. No spatial dependence of $z$ is required, allowing a unified treatment.

The hyperbolic Poincare patch (\ref{AdSmetric}) corresponds to a parameterization of the form:
\begin{eqnarray}
X_0 &=& \frac{z}{2} \left ( 1 + \frac{1 -t_h^2}{z^2} \right ) \\
X_a &=& \frac{t}{z}  \sinh(\theta) \omega_a \\
X_4 &=&  \frac{z}{2} \left ( 1 - \frac{1 +t_h^2}{z^2} \right ) \\
X_5 &=& \frac{t_h}{z}  \cosh(\theta)
\end{eqnarray}
in terms of which the metric on the hyperboloid reads
\begin{equation}
dH_3^2 = d\theta^2 + \sinh^2(\theta) d \omega_2^2.
\end{equation}
The standard AdS$_4$ slicing of AdS$_5$ still corresponds to
\begin{equation}
 \quad X_4 = \sinh(r_A) = const.
\end{equation}
which in terms of our original metric (\ref{AdSmetric}) reads
\begin{equation}
z_{AdS,1}(t_h) =  \sqrt{l^2 + t_h^2} - \sqrt{l^2-1}
\end{equation}
or
\begin{equation}
z_{AdS,2}(t_h)=\sqrt{l^2 + t_h^2} + \sqrt{l^2-1}.
\end{equation}
The two solutions correspond to an AdS$_4$ brane at positive or negative $X_4$ respectively. These are the branes located at different sides of the turning point at $X_4=0$. Either way, an AdS brane comes in from infinite $z$, reaches a minimum value of
\begin{equation}
z_{AdS}^{min} = l \mp \sqrt{l^2-1}
\end{equation}
at $t_h=0$ and moves back to infinity. It can peacefully coexist with a Minkowski brane at a fixed $z$ beyond the minimal value $z_{AdS}^{min}$. Reassuringly, the induced metric on this brane takes the form
\begin{equation}
\label{FRW}
ds^2 = - d\tau_h^2 + a(\tau_h) dH^2
\end{equation}
with
\begin{equation}
a(\tau_h) = l \sin (\tau_h/l)
\end{equation}
which is, in fact, the correct FRW parameterization of AdS$_4$.

To get the standard dS brane we use that this time
\begin{equation}
\label{desitter}
 X_0 = \cosh(r_S) = const.
\end{equation}
which translates into\footnote{A third solution can be obtained if we chose $X_0 = - \cosh(r_S)$. In this case we get $z=\sqrt{l^2+t_h^2} - \sqrt{l^2+1}$. This solution corresponds to a dS brane created at $\tau_* = - 3 \pi/2$.}
\begin{equation}
z_{dS,1}(t_h) =  \sqrt{l^2 + t_h^2} + \sqrt{l^2+1}, \quad z_{dS,2}(t_h)=\sqrt{l^2+1} - \sqrt{l^2 + t_h^2}
\end{equation}
describing the two parts of the dS brane intersecting the Poincare patch, see figure \ref{standarddesitter}
The induced metric on this brane is again of the FRW form (\ref{FRW}) with
\begin{equation}
a(\tau_h) = l \sinh (\tau_h/l)
\end{equation}
which is indeed the correct FRW parameterization of dS$_4$.

To get a more intuitive understanding of these brane embeddings, we can always switch back to standard Minkowski coordinates $x$ and $t$ on the brane. In particular, our FRW time can be expressed in terms of these as
\begin{equation} t_h^2 = t^2 - x^2
\end{equation}
where the sign ambiguity is resolved by noting that positive $t_h$ corresponds to positive $t$ and negative $t_h$ to negative $t$. It also shows clearly that the coordinate we use only cover the $t^2 \geq x^2$ part of Minkowski space.

\end{appendix}

\bibliographystyle{JHEP}
\bibliography{mismatchedbranes}

\providecommand{\href}[2]{#2}\begingroup\raggedright\begin{thebibliography}{10}

\bibitem{Randall:1999ee}
L.~Randall and R.~Sundrum, {\it {A Large mass hierarchy from a small extra
  dimension}},  {\em Phys. Rev. Lett.} {\bf 83} (1999) 3370--3373,
  \href{http://xxx.lanl.gov/abs/hep-ph/9905221}{{\tt hep-ph/9905221}}.

\bibitem{Randall:1999vf}
L.~Randall and R.~Sundrum, {\it {An Alternative to compactification}},  {\em
  Phys. Rev. Lett.} {\bf 83} (1999) 4690--4693,
  \href{http://xxx.lanl.gov/abs/hep-th/9906064}{{\tt hep-th/9906064}}.

\bibitem{Almheiri:2019hni}
A.~Almheiri, R.~Mahajan, J.~Maldacena, and Y.~Zhao, {\it {The Page curve of
  Hawking radiation from semiclassical geometry}},
  \href{http://xxx.lanl.gov/abs/1908.10996}{{\tt 1908.10996}}.

\bibitem{Rozali:2019day}
M.~Rozali, J.~Sully, M.~Van~Raamsdonk, C.~Waddell, and D.~Wakeham, {\it
  {Information radiation in BCFT models of black holes}},
  \href{http://xxx.lanl.gov/abs/1910.12836}{{\tt 1910.12836}}.

\bibitem{Karch:2000gx}
A.~Karch and L.~Randall, {\it {Open and closed string interpretation of SUSY
  CFT's on branes with boundaries}},  {\em JHEP} {\bf 06} (2001) 063,
  \href{http://xxx.lanl.gov/abs/hep-th/0105132}{{\tt hep-th/0105132}}.

\bibitem{Karch:2000ct}
A.~Karch and L.~Randall, {\it {Locally localized gravity}},  {\em JHEP} {\bf
  05} (2001) 008, \href{http://xxx.lanl.gov/abs/hep-th/0011156}{{\tt
  hep-th/0011156}}, [,140(2000)].

\bibitem{Takayanagi:2011zk}
T.~Takayanagi, {\it {Holographic Dual of BCFT}},  {\em Phys. Rev. Lett.} {\bf
  107} (2011) 101602, \href{http://xxx.lanl.gov/abs/1105.5165}{{\tt
  1105.5165}}.

\bibitem{Kachru:2003aw}
S.~Kachru, R.~Kallosh, A.~D. Linde, and S.~P. Trivedi, {\it {De Sitter vacua in
  string theory}},  {\em Phys. Rev.} {\bf D68} (2003) 046005,
  \href{http://xxx.lanl.gov/abs/hep-th/0301240}{{\tt hep-th/0301240}}.

\bibitem{Randall:2019ent}
L.~Randall, {\it {The Boundaries of KKLT}},
  \href{http://xxx.lanl.gov/abs/1912.06693}{{\tt 1912.06693}}.

\bibitem{Antonini:2019qkt}
S.~Antonini and B.~Swingle, {\it {Cosmology at the end of the world}},
  \href{http://xxx.lanl.gov/abs/1907.06667}{{\tt 1907.06667}}.

\bibitem{Banerjee:2018qey}
S.~Banerjee, U.~Danielsson, G.~Dibitetto, S.~Giri, and M.~Schillo, {\it
  {Emergent de Sitter Cosmology from Decaying Anti–de Sitter Space}},  {\em
  Phys. Rev. Lett.} {\bf 121} (2018), no.~26 261301,
  \href{http://xxx.lanl.gov/abs/1807.01570}{{\tt 1807.01570}}.

\bibitem{Banerjee:2019fzz}
S.~Banerjee, U.~Danielsson, G.~Dibitetto, S.~Giri, and M.~Schillo, {\it {de
  Sitter Cosmology on an expanding bubble}},  {\em JHEP} {\bf 10} (2019) 164,
  \href{http://xxx.lanl.gov/abs/1907.04268}{{\tt 1907.04268}}.

\bibitem{Aharony:1999ti}
O.~Aharony, S.~S. Gubser, J.~M. Maldacena, H.~Ooguri, and Y.~Oz, {\it {Large N
  field theories, string theory and gravity}},  {\em Phys. Rept.} {\bf 323}
  (2000) 183--386, \href{http://xxx.lanl.gov/abs/hep-th/9905111}{{\tt
  hep-th/9905111}}.

\bibitem{DeWolfe:1999cp}
O.~DeWolfe, D.~Z. Freedman, S.~S. Gubser, and A.~Karch, {\it {Modeling the
  fifth-dimension with scalars and gravity}},  {\em Phys. Rev.} {\bf D62}
  (2000) 046008, \href{http://xxx.lanl.gov/abs/hep-th/9909134}{{\tt
  hep-th/9909134}}.

\bibitem{Kaloper:1999sm}
N.~Kaloper, {\it {Bent domain walls as brane worlds}},  {\em Phys. Rev.} {\bf
  D60} (1999) 123506, \href{http://xxx.lanl.gov/abs/hep-th/9905210}{{\tt
  hep-th/9905210}}.

\bibitem{Kim:1999ja}
H.~B. Kim and H.~D. Kim, {\it {Inflation and gauge hierarchy in Randall-Sundrum
  compactification}},  {\em Phys. Rev.} {\bf D61} (2000) 064003,
  \href{http://xxx.lanl.gov/abs/hep-th/9909053}{{\tt hep-th/9909053}}.

\bibitem{Nihei:1999mt}
T.~Nihei, {\it {Inflation in the five-dimensional universe with an orbifold
  extra dimension}},  {\em Phys. Lett.} {\bf B465} (1999) 81--85,
  \href{http://xxx.lanl.gov/abs/hep-ph/9905487}{{\tt hep-ph/9905487}}.

\bibitem{Spradlin:2001pw}
M.~Spradlin, A.~Strominger, and A.~Volovich, {\it {Les Houches lectures on de
  Sitter space}},  in {\em {Unity from duality: Gravity, gauge theory and
  strings. Proceedings, NATO Advanced Study Institute, Euro Summer School, 76th
  session, Les Houches, France, July 30-August 31, 2001}}, pp.~423--453, 2001.
\newblock \href{http://xxx.lanl.gov/abs/hep-th/0110007}{{\tt hep-th/0110007}}.

\bibitem{Chacko:2013dra}
Z.~Chacko, R.~K. Mishra, and D.~Stolarski, {\it {Dynamics of a Stabilized
  Radion and Duality}},  {\em JHEP} {\bf 09} (2013) 121,
  \href{http://xxx.lanl.gov/abs/1304.1795}{{\tt 1304.1795}}.

\bibitem{Quiros:2019ktw}
I.~Quiros, {\it {Selected topics in scalar–tensor theories and beyond}},
  {\em Int. J. Mod. Phys.} {\bf D28} (2019), no.~07 1930012,
  \href{http://xxx.lanl.gov/abs/1901.08690}{{\tt 1901.08690}}.

\bibitem{Chacko:2001em}
Z.~Chacko and P.~J. Fox, {\it {Wave function of the radion in the dS and AdS
  brane worlds}},  {\em Phys. Rev.} {\bf D64} (2001) 024015,
  \href{http://xxx.lanl.gov/abs/hep-th/0102023}{{\tt hep-th/0102023}}.

\bibitem{Kraus:1999it}
P.~Kraus, {\it {Dynamics of anti-de Sitter domain walls}},  {\em JHEP} {\bf 12}
  (1999) 011, \href{http://xxx.lanl.gov/abs/hep-th/9910149}{{\tt
  hep-th/9910149}}.

\bibitem{Goldberger:1999uk}
W.~D. Goldberger and M.~B. Wise, {\it {Modulus stabilization with bulk
  fields}},  {\em Phys. Rev. Lett.} {\bf 83} (1999) 4922--4925,
  \href{http://xxx.lanl.gov/abs/hep-ph/9907447}{{\tt hep-ph/9907447}}.

\bibitem{Creminelli:2001th}
P.~Creminelli, A.~Nicolis, and R.~Rattazzi, {\it {Holography and the
  electroweak phase transition}},  {\em JHEP} {\bf 03} (2002) 051,
  \href{http://xxx.lanl.gov/abs/hep-th/0107141}{{\tt hep-th/0107141}}.

\end{thebibliography}\endgroup

\end{document}